\documentclass[aps,prl,twocolumn]{revtex4-1}

\usepackage{graphicx}
\usepackage[FIGTOPCAP, raggedright, nooneline]{subfigure}
\usepackage{amsmath}
\usepackage{amssymb}
\usepackage{latexsym}
\usepackage{multirow}
\usepackage{xcolor}
\usepackage[separate-uncertainty,range-phrase = --]{siunitx}
\usepackage{nicefrac}
\usepackage{tikz}

\usepackage[normalem]{ulem}
\graphicspath{{../v01/figure_files/}{../v01/}}

\begin{document}

\title{Mechanics of tissue competition: Interfaces stabilize coexistence}

\author{Nirmalendu Ganai$^{1,2}$}
\author{Tobias B\"uscher$^1$}
\author{Gerhard Gompper$^1$}
\author{Jens Elgeti$^{1,*}$}

\affiliation{$^1$Theoretical Soft Matter and Biophysics, Institute of
	Complex Systems, Forschungszentrum J\"ulich and JARA, 52425 J\"ulich, Germany}
\affiliation{$^2$Department of Physics, Nabadwip Vidyasagar College, Nabadwip, Nadia 741302, India}

\date{\today}

\begin{abstract}
Mechanical forces influence the dynamics of growing
tissues. Computer simulations are employed to study the importance of interfacial
effects in tissue competition.
It was speculated that mechanical pressure determines
the competition, where the determining quantity is the
homeostatic pressure - the pressure where division and apoptosis
balance; the tissue with the higher homeostatic pressure
overwhelms the other.
Surprisingly, a weaker tissue can persist in stable coexistence with a stronger
tissue, if adhesion between them is small enough.
An analytic continuum description can
quantitatively describe the underlying mechanism and reproduce the resulting pressures and cell-number fractions.
Computer simulations furthermore display a variety of coexisting
structures, ranging from spherical inclusions
to a bicontinuous state. 
\end{abstract}

\maketitle

%\section{TODOS}
%
%-clean up bib file\\
%-resolve ??? in text\\
%-Kuerzen. wenn finale figuren und Text
%-JSC Affiliation Jens
%\section{Introduction}
%
Mechanical forces influence the growth of cells and tissues  in several
ways \cite{Shraiman2005, Wozniak2009,Irvine2017}. This ranges from
plants adapting
their growth patterns to mechanical loads \cite{Jarvis2003,Coen2017},
all the way to tumor growth responding to mechanical forces \cite{Kumar2009,Butcher2009,Taloni2015}.
Cells have
been shown to differentiate according to substrate stiffness \cite{Engler2006},
and divide according to mechanical stress and strain \cite{Nelson2005a,Cheng2009,Fink2011,Uyttewaal2012,Streichan2014,LeGoff2015,Eder2017}.
Spheroids of many cells, grown in elastic gels \cite{Helmlinger1997,Gordon2003,Kaufman2005}
or shells \cite{Alessandri2013,Domejean2017},
or even in suspension with osmotic stress \cite{Montel2011,
  Montel2012, Delarue2013,Taloni2014}, show strong dependence of growth on
the mechanical stress from the embedding medium. 

Given the evidence of mechanical stress on growth, it seems clear that
mechanics should also influence tissue competition, such as the competition between different mutants in the imaginal wing disk of
drosophila \cite{Morata1975, Diaz2005},  or
clonal expansion in multistep cancerogenesis
\cite{Moolgavkar2003,Meza2015}.
Several theoretical studies suggested mechanics as the
underlying mechanism for both, competition \cite{Shraiman2005}
and size determination \cite{Hufnagel2007}  in the wing,
and tumor growth \cite{Basan2009, Taloni2015}.

Growth is a change of volume and the conjugate force to volume is
pressure. It stands to reason that pressure should
influence growth. A tissue grown in a finite compartment exerts a
certain pressure onto its surrounding. When reaching a steady state -
the homeostatic state - this is the homeostatic pressure $P_H$.
Under an external pressure $P$ below this value, the tissue grows;
whereas it shrinks if the pressure is above it.
This simple approach can be formulated as a linear
expansion of the bulk growth rate $k_{\text{b}}$ around the homeostatic pressure \cite{Basan2009}, 
\begin{equation}
k_{\text{b}} = \kappa (P_{\text{H}} - P)
\label{kb}
\end{equation}%
with the pressure response factor
$\kappa$. %,  the homeostatic pressure $P_{\text{H}}$ and the actual pressure $P$.
This idea has been developed  to understand
mechanical tissue comepetition in general, and metastatic inefficiency
in particular: It was argued that metastases need to reach a
critical size, below which the Laplace pressure from the
interfacial tension exeeds the homeostatic pressure difference, and
the metastasis disappears \cite{Basan2009}. 
To study the role of pressure on growth, experiments and computer simulations have been developed
to explore this effect in cell culture and in silico \cite{Montel2011,
Basan2011a, Montel2012, Delarue2013,Podewitz2015,Podewitz2016}. While confirming the
general assumption - that mechanical pressure reduces growth - these
experiments and simulations have led to another important
revelation. 
Tissues preferentially divide at the surface, even to the
extent that they die (on average) in the bulk and sustain a finite
size only by surface growth.
While consideration of nutrient transport may be neccessary for quantitative
description of some experiments \cite{Jagiella2016}, mechanics alone
already suffice.

In this work, we study the role of interfacial effects on mechanical
tissue competition by numerical simulations, in particular the effect
of  adhesive interactions between different tissues. We find that
similar to free surfaces, cells divide preferentially at the
low-adhesive interface. This interfacial growth in turn can stabilize
coexistence of two tissues with different homeostatic pressures.  
\begin{figure}
%    \begin{subfigure}
\flushleft{\hspace{.2cm}(a)\hspace{3.9cm}(b)}\\
\vspace{-0.4cm}
        \centering
          \hspace{0.5cm}
      \includegraphics[width=.4\linewidth]{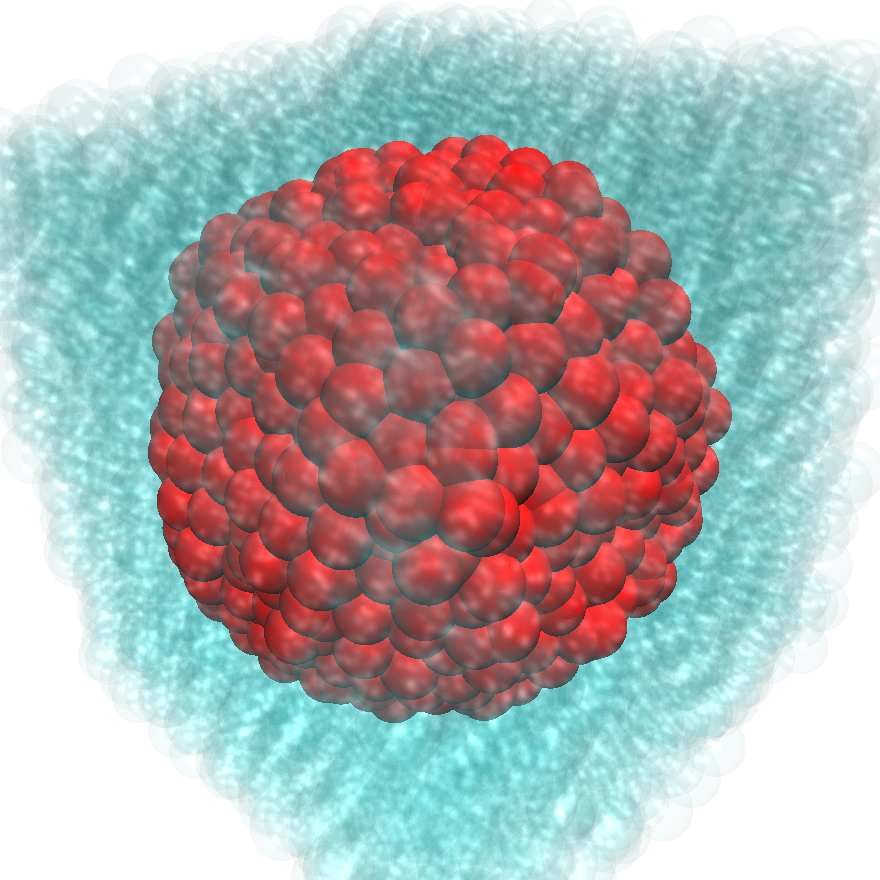}
%    \end{subfigure}
%    \begin{subfigure}
%      \centering
        \hspace{0.7cm}
        \includegraphics[width=.4\linewidth]{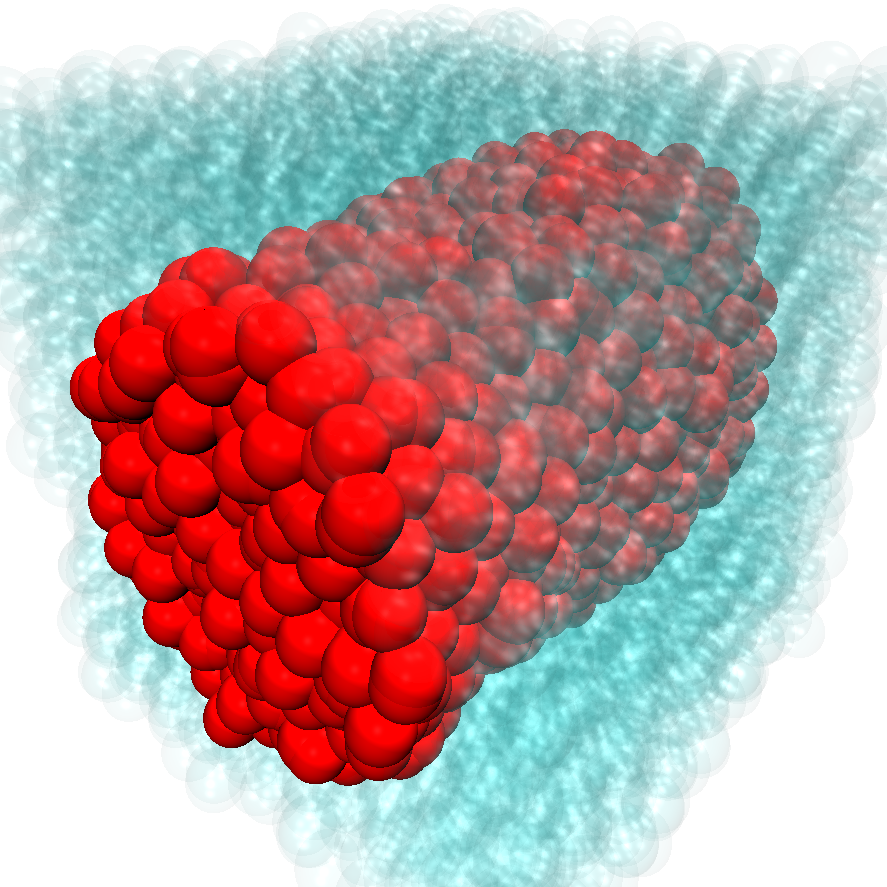}
%    \end{subfigure}
%    \begin{subfigure}
\vspace{-.7cm}
\flushleft{\hspace{.2cm}(c)\hspace{3.9cm}(d)}\\
\vspace{-0.4cm}
\centering
        \hspace{0.5cm}
        \includegraphics[width=.4\linewidth]{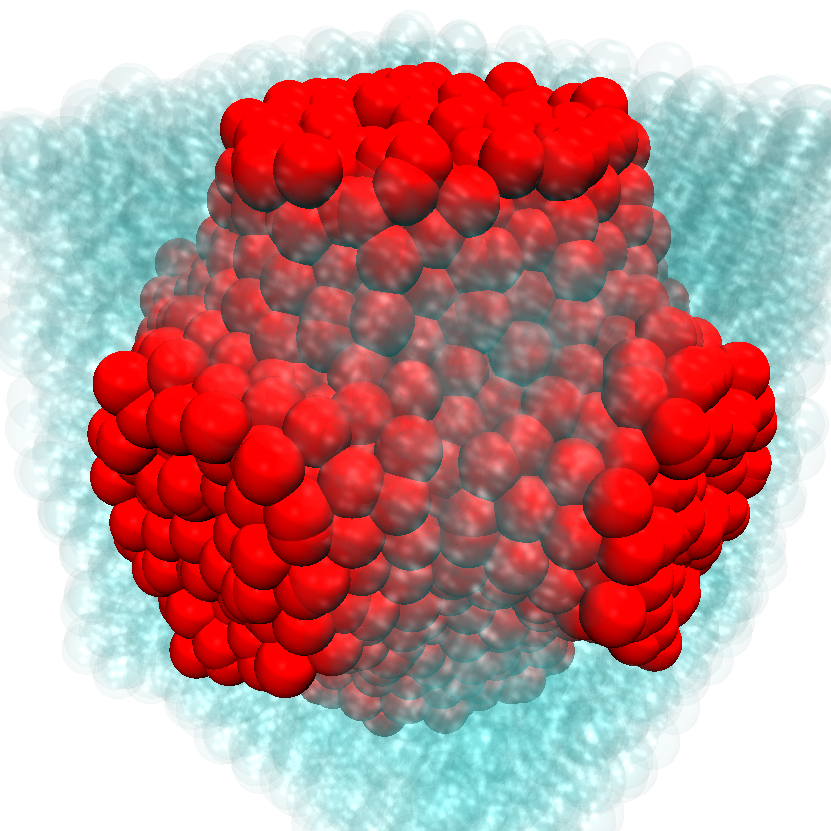}
%    \end{subfigure}
%    \begin{subfigure}
%      \centering
                \hspace{0.7cm}
        \includegraphics[width=.4\linewidth]{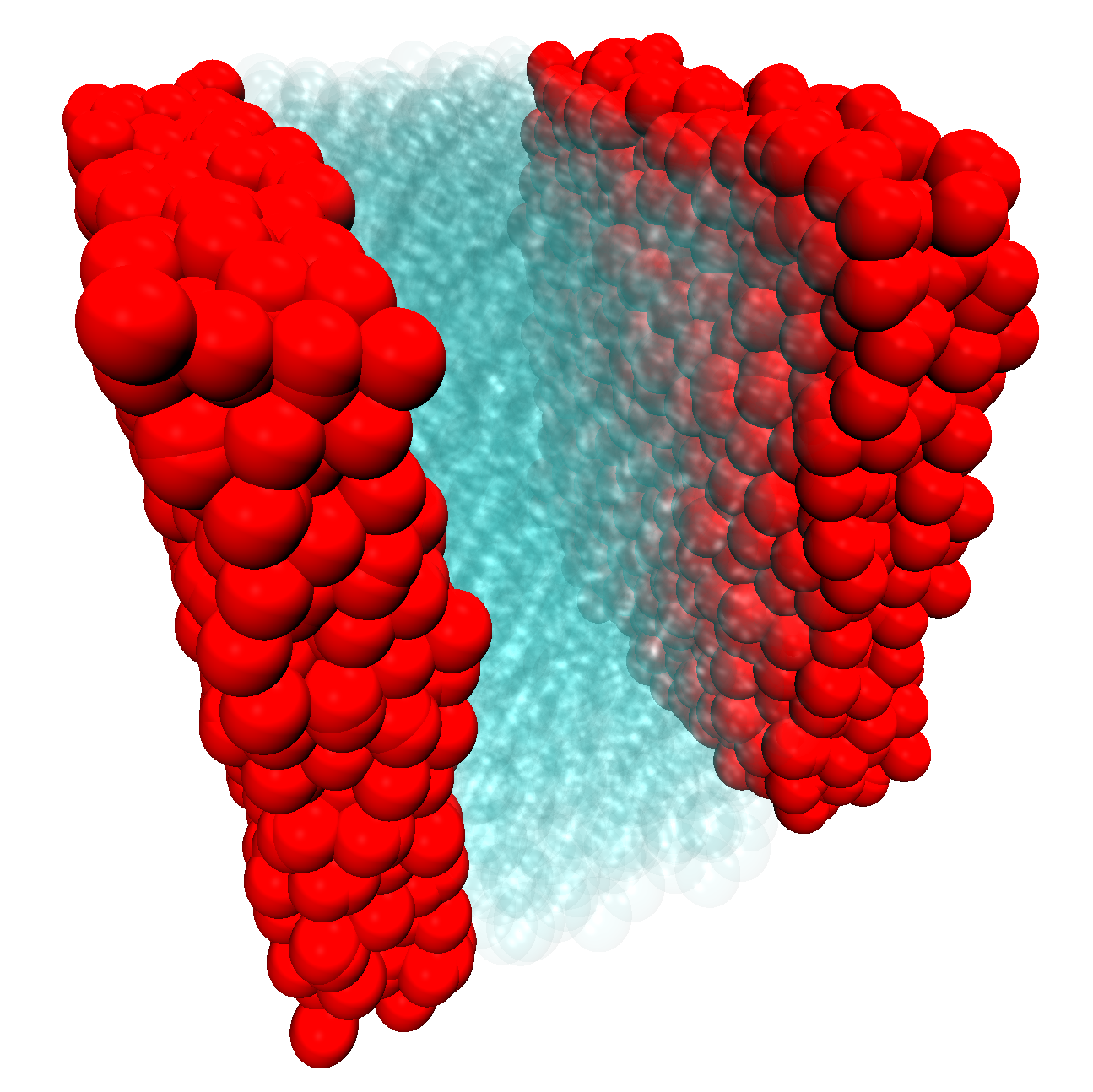}
%    \end{subfigure}

    \caption{Snapshots of various structures of tissue
        coexistence. (a) Spherical
        inclusion. (b)
        Cylindrical inclusion. (c) Schwarz-P like bicontinuous structure. (d) Flat interface. Other structures observed include
                perforated lamella, combinations (e.g. perforated lamellar together with a spheroid), and inverted (e.g. inverse spheroid) structures. }
    \label{structures}
  \end{figure}
  
Agent-based modelling of tissue growth has been very successful in the
recent years \cite{VanLiedekerke2015, Kobayashi2018}.
We follow the approach of Ref.~\cite{Basan2011a} and model growing and dividing cells by two point-like particles, which repel each other
with a growth force $	\textbf{\textit{F}}_{ij}^{\text{G}} =
\frac{G}{(r_{ij} + r_0)^2}\hat{\textbf{\textit{r}}}_{ij}$.
Once a critical distance is reached, cells divide, and two new
particles are inserted, starting the process anew. 
Apoptosis is modeled by a constant rate of cell removal $k_a$. 
Volume exclusion is maintained by a relatively soft repulsive force $\textbf{\textit{F}}_{ij}^{\text{V}}= f_0\left(\frac{R_{\text{PP}}^5}{r_{ij}^5}-1\right)\hat{\textbf{\textit{r}}}_{ij}$,
while adhesion between cells is modeled by a constant attractive
force $\textbf{\textit{F}}_{ij}^{\text{A}}= -
f_1\hat{\textbf{\textit{r}}}_{ij}$ between all cells in range
$R_{\text{PP}}$. This model results in pressure-dependent growth,
in reasonable
agreement with experiments  \cite{Montel2011,
  Basan2011a, Montel2012, Delarue2013,Podewitz2015,Podewitz2016}.
For two competing tissues A and B, parameters for each tissue can be
set independently. In this work, we only vary the growth
strength $G^{A}$ and $G^B$, the self adhesion strengths, $f_1^{\text{AA}}$,
$f_1^{\text{BB}}$ and the cross-adhesion strength $f_1^{\text{AB}}:= f_{\text{c}}$.  
See SI for further details and parameters.

To our great surprise, very small {\em cross-adhesion} strengths $f_{\text{\text{c}}}$ between cells
of different tissues (i.e. $f_{\text{c}}\ll \min(f_1^{\text{AA}}, f_1^{\text{BB}})$)
result in fundamentally different outcomes of the tissue
competition than predicted previously \cite{Basan2009}. Instead of one tissue overwhelming the other or the existence of a
critical size threshold explained above, we observe
stable coexistence in a variety of different structures depending on initial
conditions (see Fig.~\ref{structures}).
Even for two identical tissues - just without cross-adhesion - 
a single A cell in a host of B grows into a stable spheroid occupying about
a third of the volume. Similarly, a random 1:2 mixture of stronger A
cells in a host of B can result in a stable 3:1 Schwarz-P bicontiuous structure.

In order to understand this puzzling
behaviour and the underlying physical mechanisms, we turn to a simpler
initial condition of a slab-like tissue arrangement and develop an
appropriate analytic model.
\begin{figure}

          \centering
          \includegraphics[width=.8\linewidth]{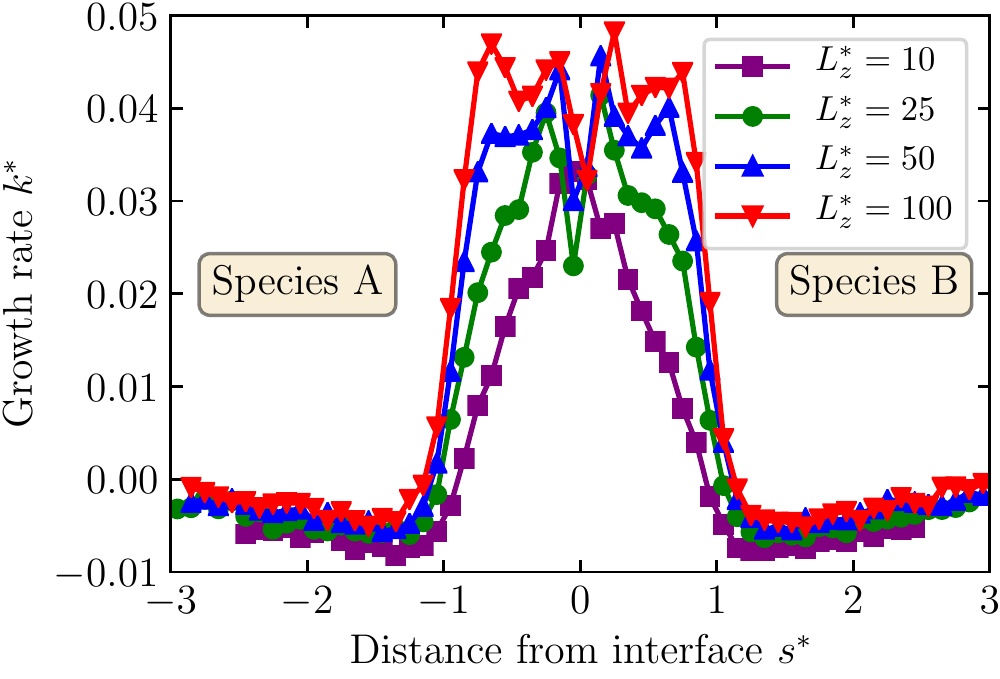}
	\caption{Growth rate $k^*$ as a function of the
		distance from the interface for the competition between two
		identical tissues with $f_{\text{c}}=0$ for different box lengths
		$L_z^*$.}
	\label{fig:schematic}
\end{figure}
%\section{Interfacial effects -- Flat interface }
Cells are confined to a finite (periodic) compartment of size
$L_x\times L_y\times L_z$. All cells in the central half ($L_z/4<z<3
L_z/4$)  are type B cells, all others type A. Large adhesion between
cells of the same tissue and no adhesion between cells of different
tissues leads to a large surface tension, stabilizing the flat
interface.
The division profile (see Fig. \ref{fig:schematic}) reveals that cells
divide more in a small region of width $a$ (about one or two cell layers) at the interface. In the
bulk of the tissue, the net growth is negative due to an elevated
pressure. These results motivate a two-rate growth model \cite{Montel2011,
  Basan2011a, Montel2012, Delarue2013, Podewitz2015}
\begin{align}
	\partial_t \rho +\nabla\cdot (\rho \textbf{\textit{v}}) =k_{\text{b}}\rho +\Delta k_{\text{s}}
  \Theta(s-a) \rho\text{,}
\end{align}
where $\rho$ is the cellular density, $\Theta$ the Heavyside step
function, $s$ the distance to the nearest interface and $\textbf{\textit{v}}$ the cell-velocity field. 
 The additional growth at the interface is modeled as a growth enhancement $\Delta k_{\text{s}}$ near the interface (less than $a$
 away).

\begin{figure}

		\centering
		\includegraphics[width=.8\linewidth]{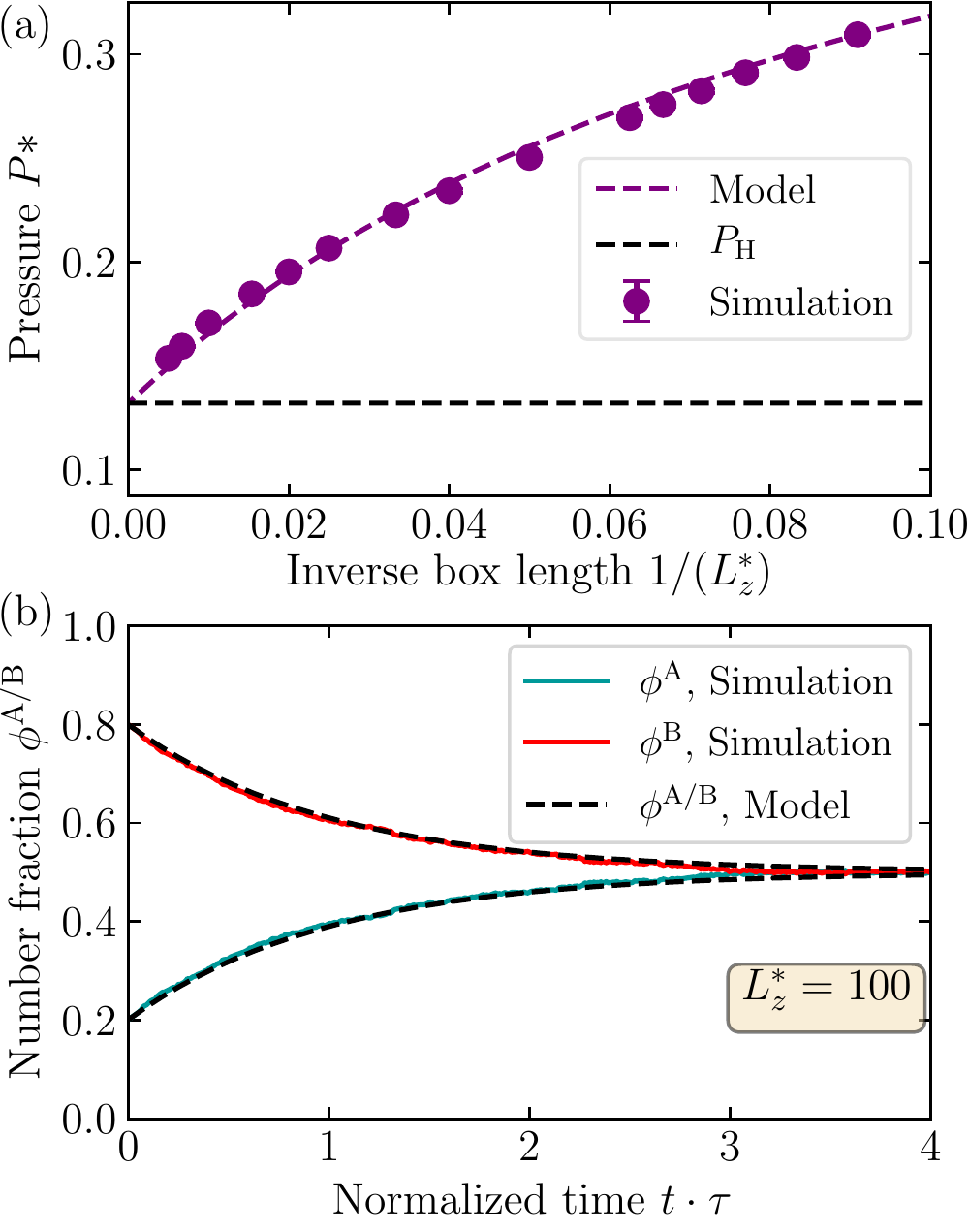}

	\caption{(a) Average pressure measured in competitions between
          two identical tissue (reference tissue, see SI) with zero cross-adhesion $f_{\text{c}}=0$ in terms of the inverse box length $L_z^*$. Dashed purple line shows the prediction of the two-rate model according to Eq. \eqref{pressure_same}. (b) Solid cyan and red lines show the time evolution of the cell number fractions $\phi^{\text{A/B}}$ for a competition as in (a), for a box length $L_z^* = 100$. Dashed black line shows the prediction given by Eq. \eqref{dyn} for both tissues.}
	\label{same}
\end{figure}

 Division and apoptosis events locally relax stress and thus lead to
 a liquidification of the tissue on longer timescales \cite{Ranft2010,   Khalilgharibi2016,Matoz-Fernandez2017a}.
 Indeed, experiments on tissue rheology suggest liquid behaviour on
 long timescales \cite{Phillips1978,Guevorkian2010, Gonzalez-Rodriguez2013},
while some experiments on drosophila wing discs suggest that not all
stress is relaxed by 
 growth \cite{LeGoff2013,Mao2013,Pan2016}.
Our model tissue clearly behaves as a liquid
\cite{Ranft2010}. With the low velocities and no external forcing, we can thus assume a constant pressure across
the system. This motivates expanding $k_\text{b}$ as in
Eq.~\eqref{kb}, and similarly $\Delta k_{\text{s}} \simeq \Delta k_{\text{s}}^0 + \Delta
k_{\text{s}}^1(P_\text{H}-P)$. 
Under the assumption of constant density and with an integration over the system, 
the time evolution of the cell number fraction
$\phi=N_{\text{A}}/(N_{\text{A}}+N_{\text{B}})$  of type A cells reads
\begin{align}
	\partial_t \phi =k_{\text{b}}\phi +\Delta k_{\text{s}} \phi_{\text{s}},\label{interface}
\end{align}
with the fraction $\phi_{\text{s}}$ of A type cells at the surface.
Two identical tissues
(without cross-adhesion) then develop two interfaces $L_z/2$ apart.
Insertion of the linear expansions in Eq. \eqref{interface} then
yields the pressure 
\begin{equation}
	P = P_{\text{H}} + \frac {4a \Delta k_{\text{s}}^{0}}{(4a \Delta k_{\text{s}}^{1} + \kappa
		L_z)}\text{,}\label{pressure_same}
            \end{equation}
i.e. the additional growth at the interface elevates the pressure
above the homeostatic pressure, which in turn causes the negative net growth rate in the
bulk. We determine the bulk parameters $P_{\text{H}}, \kappa$ from
bulk simulations as in Ref.~\cite{Podewitz2015}, and the surface
parameters $a \Delta k_{\text{s}}^{0}, a \Delta k_{\text{s}}^{1} $ by
fitting Eq.~\eqref{pressure_same} to a tissue with mirror boundary
conditions in one direction (see SI). As shown in
Ref.~\cite{Podewitz2015}, the homeostatic pressure grows approximately
linearly with $G$, and decreases linearly with $f_1$. $\kappa$ is essentially
independent off $f_1$, but decreases linearly with $G$. The surface parameter $\Delta k_{\text{s}}^0$ is only weakly
dependent on $G$, but grows linearly with $f_1$, while $\Delta k_{\text{s}}^1$ does not show a clear dependence on tissue parameter (see SI).
Representatively, we show the pressure dependence on box length $L_z$
for two identical tissues without cross adhesion. 
With the parameters fixed, the
theory reproduces the 
simulations without further parameter adjustment (see Fig.~ \ref{same}(a)).

For identical tissues, the steady-state solution % ($\partial_t\phi=0$)
to Eq. \eqref{interface} is $\phi=1/2$ by symmetry.
For the dynamics we obtain
\begin{equation}
	\phi(t) = \frac{1}{2} + (\phi_0 - \frac{1}{2})e^{-\kappa(P-P_{\text{H}})t}\text{,}\label{dyn}
\end{equation}
with the initial number fraction $\phi_0$.
As shown in  Fig. \ref{same}(b), Eq. \eqref{dyn}
reproduces the simulation dynamics.

\begin{figure}

		\centering
		\includegraphics[width=.8\linewidth]{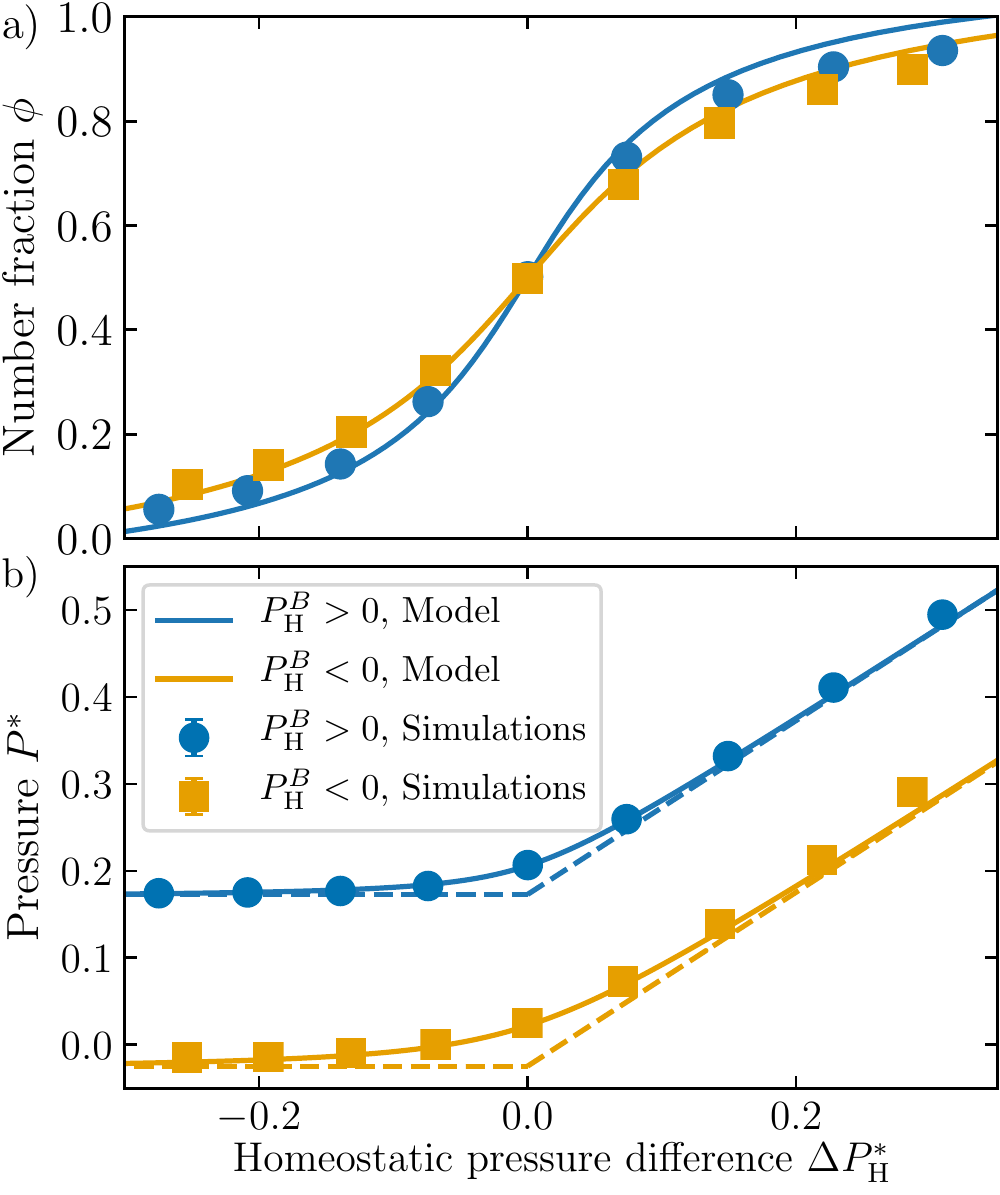}\\

	\caption{(a) Cell number fractions $\phi$ for various
		homeostatic pressure differences $P_{\text{H}}^{B} -
		P_{\text{H}}^{A}$. % when tissue B is in competition with a
		% non-identical tissue A via
		% $f_{\text{c}}=0$
		%in a box of $L_z=40$.
                Tissue B is fixed (reference tissue) and the
                homeostatic pressure of tissue A is varied. Symbols
                are simulation results while the solid lines are
                predictions by the two-rate model according to
                Eq.~\eqref{different}, using the parameters of tissue
                B.  Blue corresponds to positive homeostatic pressure
                of tissue B and yellow to a negative one. (b) Average
                pressure measured during the simulations shown in (a)
                together with a plot of
                Eq. \eqref{pressure_different}, using the parameter of
                tissue B. Dashed lines are lower bounds from
                $L_{A,B}<L_z$.  Boxsize $L_x^*=L_y^*=10$;$L_z^*=40$ }
	\label{simth2}
      \end{figure}

Next, we explore the competition between two different tissues with a planar
interface. We balance the pressures on both sides of the interface and
get
\begin{equation}
	P = P_{\text{H}}^A + \frac {2a \Delta k_{\text{s}}^{0A}}{(2a \Delta k_{\text{s}}^{1A} + \kappa^A L_A)} =  P_{\text{H}}^B + \frac {2a \Delta k_{\text{s}}^{0B}}{(2a \Delta k_{\text{s}}^{1B} + \kappa^B L_B)},
	\label{pressure_different}
\end{equation}
where $L_B$ and $L_A(=L_z -L_B)$ are the lengths occupied by tissue
A and B.
Note that inserting $L_{A,B}<L_z$ in Eq.~\eqref{pressure_different}
gives a lower bound for the pressure: The system pressure is always
larger than the homeostatic pressure of the stronger tissue, plus a
system-size-dependent constant. Indeed, this lower bound describes the
pressure rather well. The stronger tissue occupies the larger part of
the system, and thus $L_{A,B}\approx L_z$. Thus the pressure is almost
constant for $\Delta P_H <0$, and grows almost linearly for $\Delta P_H >
0$ (see Fig.~\ref{simth2}).  The weaker tissue suports the higher
pressure by decreasing in size, and thus its apoptotic volume,
sustained by surface growth.
For the simulated tissues, the parameter
$\kappa$, $\Delta k_{\text{s}}^0$ and $\Delta k_{\text{s}}^1$ only
show small variations (see SI). We therefore assume them to be the same for both tissues in order to obtain
\begin{align}
	\begin{split}
		\phi &= \frac{1}{2}+\frac{2a\Delta k_{\text{s}}^{0}}{\kappa (P_{\text{H}}^{{B}}-P_{\text{H}}^{{A}})L_z} \\
		&\pm \Bigg[ \bigg(\frac{2a\Delta k_{\text{s}}^{0}}{\kappa (P_{\text{H}}^{{B}}-P_{\text{H}}^{{A}})L_z} \bigg)^2 + \bigg(\frac{1}{2} + \frac{2a\Delta k_{\text{s}}^{1}}{\kappa L_z} \bigg)^2 \Bigg]^{\frac{1}{2}}
	\text{.}\end{split}\label{different}
\end{align}
Note that for $\Delta P_{\text{H}} \equiv
(P_{\text{H}}^{{B}}-P_{\text{H}}^{{A}})\to 0$, Eq. \eqref{different}
reproduces $\phi=1/2$ as expected. Around $\Delta P_{\text{H}} = 0$,
$\phi$ grows linearly with  $\Delta P_{\text{H}}$ and then slows down (see Fig.~\ref{simth2}). For large differences in homeostatic pressure, the model predicts two interfaces less than $2a$ apart, thus violating its assumptions, and consequently fails to predict the simulation results properly.
Equations~\eqref{pressure_different} and \eqref{different} are able to reproduce simulation results fairly
well without parameter adjustments (see Fig.~\ref{simth2}) in a broad
parameter regime. 
Note that this also holds true for negative homeostatic
bulk pressures, where indeed the system pressure is positive,
thanks to the surface growth (see Eq.~\eqref{pressure_different}).

%%%%%%%%%%%%%%%%%%%%%%%%%%%%%%%%%%%%%%%%%%%%%%%%%%%%%%      
%\section{Non-planar interfaces}

\begin{figure}

	\centering
	\includegraphics[width=0.8\linewidth]{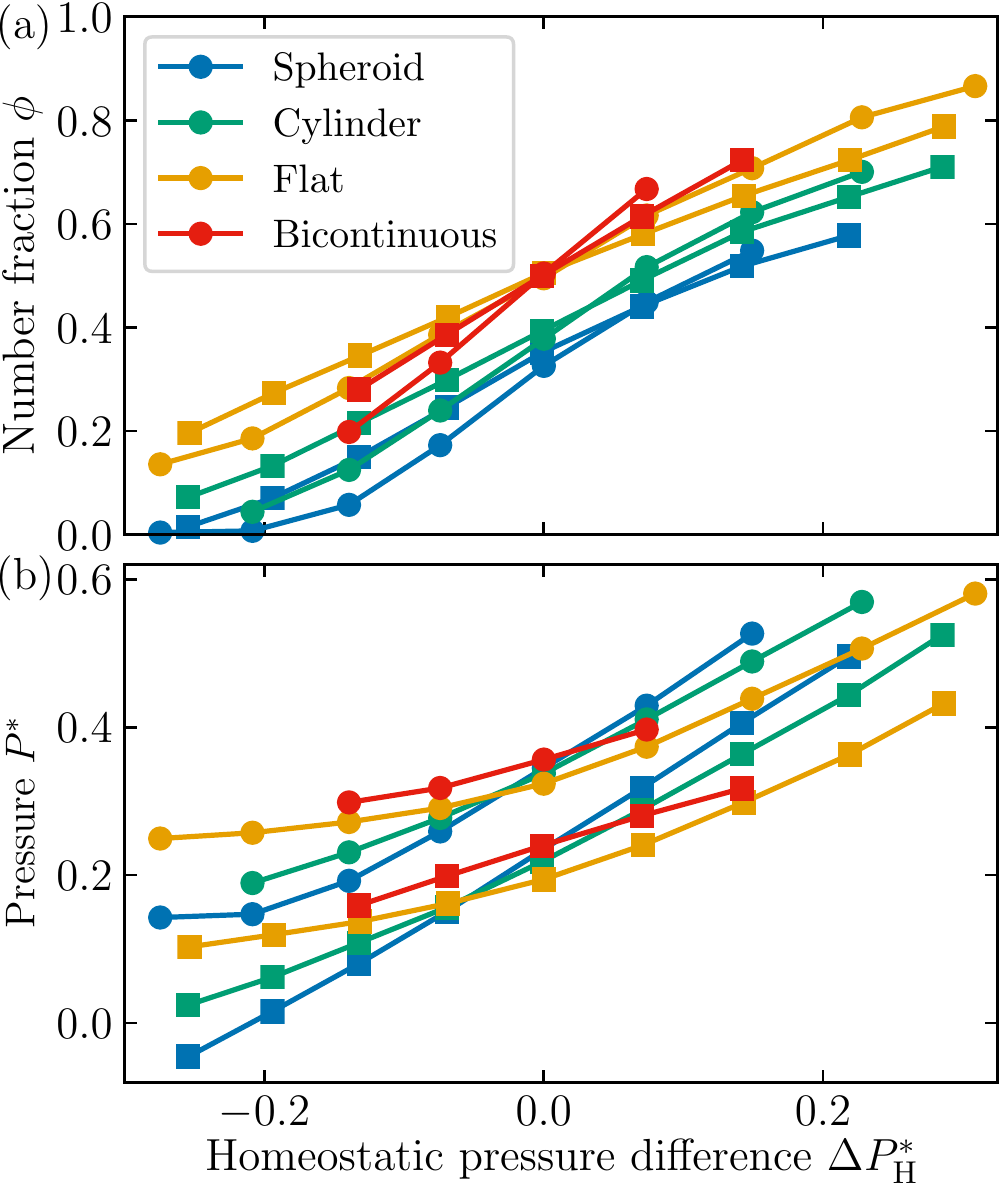}
	\caption{Cell number fractions  $\phi$ for different
		homeostatic pressure differences $\Delta P_{\text{H}}^*$ and
		different structures, as indicated by color. %namely a spherical (blue color)
                %and a cylindrical %???Shorten: as indicated by color?%%%
		%inclusion (green) of tissue A in B, a bicontinuous phase (red) and for comparison a flat
		%interface (yellow).
                Circles correspond to a positive homeostatic
		pressure of tissues B and squares to a negative one (same
		parameters as in Fig. \ref{simth2}, except cubic box size $L=10$). (b) Average pressure measured in the simulations shown in (a)}
	\label{cylinder_spheroid}
      \end{figure}

These results show that indeed the enhanced growth at the interface
lies at the heart of the coexistence of tissues observed in our simulations.
However, a flat interface is not the only stable structure for two
competing tissues. Depending on initial conditions and parameters, a large range of
other structures can be found (see Fig.~\ref{structures}).
These different structures result in different surface-to-volume
ratios (and possibly other interfacial effects), changing the steady-state volume fractions and pressures. 
We present simulation results for these structures in Fig. \ref{cylinder_spheroid}.

Compared to flat interfaces, the number fraction $\phi$ of tissues in
spherical or cylindrical configuration is smaller, with spheroids
being smaller than cylinders.
Note that spheroids become unstable with growing homeostatic pressure difference. They then turn into cylinders, which again become unstable
and turn into a slab-like structure, which probably becomes unstable
as well. Vice versa, cylinders turn into spheroids if the difference
in homeostatic pressure is very negative. The number fraction of the
bicontinuous phase is roughly the same as for flat interfaces, but the
bicontinuous phase is only stable in a small regime of homeostatic
pressure differences.
For larger differences in homeostatic pressure  it turns into a
perforated lamella phase of the weaker tissue inside the stronger
tissue.
In general, the number fraction $\phi$ of all structures changes
sigmoidally with homeostatic pressure difference (see Fig.~\ref{structures}).
\begin{figure}
	\centering
	\includegraphics[width=0.8\linewidth]{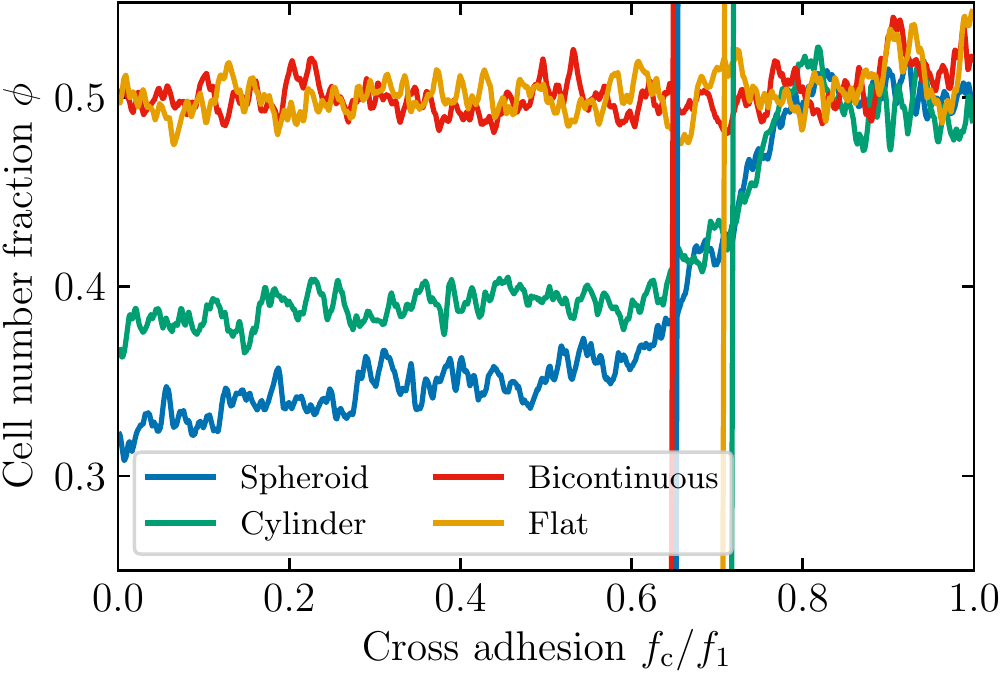}
	\caption{Variation of cell number fraction $\phi$ with 
		time with increasing cross-adhesion $f_{\text{c}}/f_1=t^*/240$ between two identical
		tissues. Simulations are started from spherical (blue)
		and cylindrical inclusions(green)
		of tissue A in B as well as from flat interfaces (yellow) and a bicontinuous phase (red). Solid lines are marking
		transition points after which the corresponding initial
		structure forms a three dimensional percolated
		cluster. Cubic box size $L=10$}
	\label{increase_fc}
\end{figure}

While all of these structures are very stable over time, the
question arises how stable they are when the interfacial effects
become smaller. We study this effect numerically, by observing the
structures for two identical tissues formed under zero cross-adhesion
and continuously increase the cross-adhesion $f_c$ to the value of
self-adhesion (i.e. $f_c=f_1^{AA}=f_1^{BB}$). Figure \ref{increase_fc}
shows that all structures remain almost unchanged up to a
cross-adhesion $f_{\text{c}}$ approximately two thirds of the self
adhesion $f_1$. For higher  $f_{\text{c}}$ %cross-adhesions 
only a mixed, sponge-like state remains.

%\section{conclusions}
In summary, the interface between two tissues plays
an important role in the competition between them. The enhanced growth
at the interface can stabilize coexisting phases even when one tissue
has a higher homeostatic pressure.
The coexisting phase appears in a variety of different structures,
ranging from a spherical inclusion over a flat interface to a
bicontinous phase.

Interesting future directions are interfacial dynamics, roughness, and shapes, as previously explored
for tissues on substrates and without additional interfacial growth
\cite{Ranft2014,Podewitz2016,Williamson2017}. Vice versa, it would be
interesting to add interfacial growth to tissues growing on substrates.

Finally, our results suggests a tentative
explanation for tumor heterogeneity and the abundance of occult tumors: small symptom-free
micro-tumors that are frequently found in the human body
\cite{Bissell2011}.
%First reported in 1935 for the prostate \cite{Rich1935,Rich2007},
%it has now been frequently reported \cite{Sakr1993}.
For the
theyroid, it is indeed considered 'normal' to find microscopic lesions
\cite{Harach1985}. Our results provide a mechanical explanation how
coexsistence of different tissues can be stable by simple mechanical
effects. For example, a mutation might downregulate cadherins - an
important cellular adhesion protein -  as it
often happens in tumors \cite{weinbergbook}. On the one hand, this
might reduce survival signaling \cite{molecularbio}, but the lack of
adhesion also favors our mechanism of coexistence, even for weaker
tissue growth.

\section{Acknowledgements}
The authors gratefully acknowledge the computing time granted through JARA-HPC on the supercomputer JURECA \cite{jureca} at Forschungszentrum Jülich
\bibliography{all.bib}

%merlin.mbs apsrev4-1.bst 2010-07-25 4.21a (PWD, AO, DPC) hacked
%Control: key (0)
%Control: author (8) initials jnrlst
%Control: editor formatted (1) identically to author
%Control: production of article title (-1) disabled
%Control: page (0) single
%Control: year (1) truncated
%Control: production of eprint (0) enabled
\begin{thebibliography}{53}%
\makeatletter
\providecommand \@ifxundefined [1]{%
 \@ifx{#1\undefined}
}%
\providecommand \@ifnum [1]{%
 \ifnum #1\expandafter \@firstoftwo
 \else \expandafter \@secondoftwo
 \fi
}%
\providecommand \@ifx [1]{%
 \ifx #1\expandafter \@firstoftwo
 \else \expandafter \@secondoftwo
 \fi
}%
\providecommand \natexlab [1]{#1}%
\providecommand \enquote  [1]{``#1''}%
\providecommand \bibnamefont  [1]{#1}%
\providecommand \bibfnamefont [1]{#1}%
\providecommand \citenamefont [1]{#1}%
\providecommand \href@noop [0]{\@secondoftwo}%
\providecommand \href [0]{\begingroup \@sanitize@url \@href}%
\providecommand \@href[1]{\@@startlink{#1}\@@href}%
\providecommand \@@href[1]{\endgroup#1\@@endlink}%
\providecommand \@sanitize@url [0]{\catcode `\\12\catcode `\$12\catcode
  `\&12\catcode `\#12\catcode `\^12\catcode `\_12\catcode `\%12\relax}%
\providecommand \@@startlink[1]{}%
\providecommand \@@endlink[0]{}%
\providecommand \url  [0]{\begingroup\@sanitize@url \@url }%
\providecommand \@url [1]{\endgroup\@href {#1}{\urlprefix }}%
\providecommand \urlprefix  [0]{URL }%
\providecommand \Eprint [0]{\href }%
\providecommand \doibase [0]{http://dx.doi.org/}%
\providecommand \selectlanguage [0]{\@gobble}%
\providecommand \bibinfo  [0]{\@secondoftwo}%
\providecommand \bibfield  [0]{\@secondoftwo}%
\providecommand \translation [1]{[#1]}%
\providecommand \BibitemOpen [0]{}%
\providecommand \bibitemStop [0]{}%
\providecommand \bibitemNoStop [0]{.\EOS\space}%
\providecommand \EOS [0]{\spacefactor3000\relax}%
\providecommand \BibitemShut  [1]{\csname bibitem#1\endcsname}%
\let\auto@bib@innerbib\@empty
%</preamble>
\bibitem [{\citenamefont {Shraiman}(2005)}]{Shraiman2005}%
  \BibitemOpen
  \bibfield  {author} {\bibinfo {author} {\bibfnamefont {B.~I.}\ \bibnamefont
  {Shraiman}},\ }\href@noop {} {\bibfield  {journal} {\bibinfo  {journal}
  {Proc. Natl. Acad. Sci. U. S. A.}\ }\textbf {\bibinfo {volume} {102}},\
  \bibinfo {pages} {3318} (\bibinfo {year} {2005})}\BibitemShut {NoStop}%
\bibitem [{\citenamefont {Wozniak}\ and\ \citenamefont
  {Chen}(2009)}]{Wozniak2009}%
  \BibitemOpen
  \bibfield  {author} {\bibinfo {author} {\bibfnamefont {M.~A.}\ \bibnamefont
  {Wozniak}}\ and\ \bibinfo {author} {\bibfnamefont {C.~S.}\ \bibnamefont
  {Chen}},\ }\href@noop {} {\bibfield  {journal} {\bibinfo  {journal} {Nat.
  Rev. Mol. Cell Biol.}\ }\textbf {\bibinfo {volume} {10}},\ \bibinfo {pages}
  {34} (\bibinfo {year} {2009})}\BibitemShut {NoStop}%
\bibitem [{\citenamefont {Irvine}\ and\ \citenamefont
  {Shraiman}(2017)}]{Irvine2017}%
  \BibitemOpen
  \bibfield  {author} {\bibinfo {author} {\bibfnamefont {K.~D.}\ \bibnamefont
  {Irvine}}\ and\ \bibinfo {author} {\bibfnamefont {B.~I.}\ \bibnamefont
  {Shraiman}},\ }\href@noop {} {\bibfield  {journal} {\bibinfo  {journal}
  {Development}\ }\textbf {\bibinfo {volume} {144}},\ \bibinfo {pages} {4238}
  (\bibinfo {year} {2017})}\BibitemShut {NoStop}%
\bibitem [{\citenamefont {Jarvis}\ \emph {et~al.}(2003)\citenamefont {Jarvis},
  \citenamefont {Briggs},\ and\ \citenamefont {Knox}}]{Jarvis2003}%
  \BibitemOpen
  \bibfield  {author} {\bibinfo {author} {\bibfnamefont {M.}~\bibnamefont
  {Jarvis}}, \bibinfo {author} {\bibfnamefont {S.}~\bibnamefont {Briggs}}, \
  and\ \bibinfo {author} {\bibfnamefont {J.}~\bibnamefont {Knox}},\ }\href@noop
  {} {\bibfield  {journal} {\bibinfo  {journal} {Plant Cell Environ.}\ }\textbf
  {\bibinfo {volume} {{26}}},\ \bibinfo {pages} {{977}} (\bibinfo {year}
  {{2003}})}\BibitemShut {NoStop}%
\bibitem [{\citenamefont {Coen}\ \emph {et~al.}(2017)\citenamefont {Coen},
  \citenamefont {Kennaway},\ and\ \citenamefont {Whitewoods}}]{Coen2017}%
  \BibitemOpen
  \bibfield  {author} {\bibinfo {author} {\bibfnamefont {E.}~\bibnamefont
  {Coen}}, \bibinfo {author} {\bibfnamefont {R.}~\bibnamefont {Kennaway}}, \
  and\ \bibinfo {author} {\bibfnamefont {C.}~\bibnamefont {Whitewoods}},\
  }\href@noop {} {\bibfield  {journal} {\bibinfo  {journal} {Development}\
  }\textbf {\bibinfo {volume} {144}},\ \bibinfo {pages} {4203} (\bibinfo {year}
  {2017})}\BibitemShut {NoStop}%
\bibitem [{\citenamefont {Kumar}\ and\ \citenamefont
  {Weaver}(2009)}]{Kumar2009}%
  \BibitemOpen
  \bibfield  {author} {\bibinfo {author} {\bibfnamefont {S.}~\bibnamefont
  {Kumar}}\ and\ \bibinfo {author} {\bibfnamefont {V.~M.}\ \bibnamefont
  {Weaver}},\ }\href@noop {} {\bibfield  {journal} {\bibinfo  {journal} {Cancer
  Metastasis Rev.}\ }\textbf {\bibinfo {volume} {{28}}},\ \bibinfo {pages}
  {{113}} (\bibinfo {year} {{2009}})}\BibitemShut {NoStop}%
\bibitem [{\citenamefont {Butcher}\ \emph {et~al.}(2009)\citenamefont
  {Butcher}, \citenamefont {Alliston},\ and\ \citenamefont
  {Weaver}}]{Butcher2009}%
  \BibitemOpen
  \bibfield  {author} {\bibinfo {author} {\bibfnamefont {D.~T.}\ \bibnamefont
  {Butcher}}, \bibinfo {author} {\bibfnamefont {T.}~\bibnamefont {Alliston}}, \
  and\ \bibinfo {author} {\bibfnamefont {V.~M.}\ \bibnamefont {Weaver}},\
  }\href@noop {} {\bibfield  {journal} {\bibinfo  {journal} {Nat. Rev. Cancer}\
  }\textbf {\bibinfo {volume} {9}},\ \bibinfo {pages} {108} (\bibinfo {year}
  {2009})}\BibitemShut {NoStop}%
\bibitem [{\citenamefont {Taloni}\ \emph {et~al.}(2015)\citenamefont {Taloni},
  \citenamefont {Ben~Amar}, \citenamefont {Zapperi},\ and\ \citenamefont
  {La~Porta}}]{Taloni2015}%
  \BibitemOpen
  \bibfield  {author} {\bibinfo {author} {\bibfnamefont {A.}~\bibnamefont
  {Taloni}}, \bibinfo {author} {\bibfnamefont {M.}~\bibnamefont {Ben~Amar}},
  \bibinfo {author} {\bibfnamefont {S.}~\bibnamefont {Zapperi}}, \ and\
  \bibinfo {author} {\bibfnamefont {C.~A.}\ \bibnamefont {La~Porta}},\
  }\href@noop {} {\bibfield  {journal} {\bibinfo  {journal} {Eur. Phys. J.
  Plus}\ }\textbf {\bibinfo {volume} {130}},\ \bibinfo {pages} {224} (\bibinfo
  {year} {2015})}\BibitemShut {NoStop}%
\bibitem [{\citenamefont {Engler}\ \emph {et~al.}(2006)\citenamefont {Engler},
  \citenamefont {Sen}, \citenamefont {Sweeney},\ and\ \citenamefont
  {Discher}}]{Engler2006}%
  \BibitemOpen
  \bibfield  {author} {\bibinfo {author} {\bibfnamefont {A.~J.}\ \bibnamefont
  {Engler}}, \bibinfo {author} {\bibfnamefont {S.}~\bibnamefont {Sen}},
  \bibinfo {author} {\bibfnamefont {H.~L.}\ \bibnamefont {Sweeney}}, \ and\
  \bibinfo {author} {\bibfnamefont {D.~E.}\ \bibnamefont {Discher}},\
  }\href@noop {} {\bibfield  {journal} {\bibinfo  {journal} {Cell}\ }\textbf
  {\bibinfo {volume} {126}},\ \bibinfo {pages} {677 } (\bibinfo {year}
  {2006})}\BibitemShut {NoStop}%
\bibitem [{\citenamefont {Nelson}\ \emph {et~al.}(2005)\citenamefont {Nelson},
  \citenamefont {Jean}, \citenamefont {Tan}, \citenamefont {Liu}, \citenamefont
  {Sniadecki}, \citenamefont {Spector},\ and\ \citenamefont
  {Chen}}]{Nelson2005a}%
  \BibitemOpen
  \bibfield  {author} {\bibinfo {author} {\bibfnamefont {C.~M.}\ \bibnamefont
  {Nelson}}, \bibinfo {author} {\bibfnamefont {R.~P.}\ \bibnamefont {Jean}},
  \bibinfo {author} {\bibfnamefont {J.~L.}\ \bibnamefont {Tan}}, \bibinfo
  {author} {\bibfnamefont {W.~F.}\ \bibnamefont {Liu}}, \bibinfo {author}
  {\bibfnamefont {N.~J.}\ \bibnamefont {Sniadecki}}, \bibinfo {author}
  {\bibfnamefont {A.~A.}\ \bibnamefont {Spector}}, \ and\ \bibinfo {author}
  {\bibfnamefont {C.~S.}\ \bibnamefont {Chen}},\ }\href@noop {} {\bibfield
  {journal} {\bibinfo  {journal} {Proc. Natl. Acad. Sci. U.S.A.}\ }\textbf
  {\bibinfo {volume} {102}},\ \bibinfo {pages} {11594} (\bibinfo {year}
  {2005})}\BibitemShut {NoStop}%
\bibitem [{\citenamefont {Cheng}\ \emph {et~al.}(2009)\citenamefont {Cheng},
  \citenamefont {Tse}, \citenamefont {Jain},\ and\ \citenamefont
  {Munn}}]{Cheng2009}%
  \BibitemOpen
  \bibfield  {author} {\bibinfo {author} {\bibfnamefont {G.}~\bibnamefont
  {Cheng}}, \bibinfo {author} {\bibfnamefont {J.}~\bibnamefont {Tse}}, \bibinfo
  {author} {\bibfnamefont {R.~K.}\ \bibnamefont {Jain}}, \ and\ \bibinfo
  {author} {\bibfnamefont {L.~L.}\ \bibnamefont {Munn}},\ }\href@noop {}
  {\bibfield  {journal} {\bibinfo  {journal} {PLoS One}\ }\textbf {\bibinfo
  {volume} {4}},\ \bibinfo {pages} {e4632} (\bibinfo {year}
  {2009})}\BibitemShut {NoStop}%
\bibitem [{\citenamefont {Fink}\ \emph {et~al.}(2011)\citenamefont {Fink},
  \citenamefont {Carpi}, \citenamefont {Betz}, \citenamefont {Betard},
  \citenamefont {Chebah}, \citenamefont {Azioune}, \citenamefont {Bornens},
  \citenamefont {Sykes}, \citenamefont {Fetler}, \citenamefont {Cuvelier},\
  and\ \citenamefont {Piel}}]{Fink2011}%
  \BibitemOpen
  \bibfield  {author} {\bibinfo {author} {\bibfnamefont {J.}~\bibnamefont
  {Fink}}, \bibinfo {author} {\bibfnamefont {N.}~\bibnamefont {Carpi}},
  \bibinfo {author} {\bibfnamefont {T.}~\bibnamefont {Betz}}, \bibinfo {author}
  {\bibfnamefont {A.}~\bibnamefont {Betard}}, \bibinfo {author} {\bibfnamefont
  {M.}~\bibnamefont {Chebah}}, \bibinfo {author} {\bibfnamefont
  {A.}~\bibnamefont {Azioune}}, \bibinfo {author} {\bibfnamefont
  {M.}~\bibnamefont {Bornens}}, \bibinfo {author} {\bibfnamefont
  {C.}~\bibnamefont {Sykes}}, \bibinfo {author} {\bibfnamefont
  {L.}~\bibnamefont {Fetler}}, \bibinfo {author} {\bibfnamefont
  {D.}~\bibnamefont {Cuvelier}}, \ and\ \bibinfo {author} {\bibfnamefont
  {M.}~\bibnamefont {Piel}},\ }\href@noop {} {\bibfield  {journal} {\bibinfo
  {journal} {Nat. Cell Biol.}\ }\textbf {\bibinfo {volume} {13}},\ \bibinfo
  {pages} {771} (\bibinfo {year} {2011})}\BibitemShut {NoStop}%
\bibitem [{\citenamefont {Uyttewaal}\ \emph {et~al.}(2012)\citenamefont
  {Uyttewaal}, \citenamefont {Burian}, \citenamefont {Alim}, \citenamefont
  {Landrein}, \citenamefont {Borowska-Wykret}, \citenamefont {Dedieu},
  \citenamefont {Peaucelle}, \citenamefont {Ludynia}, \citenamefont {Traas},
  \citenamefont {Boudaoud}, \citenamefont {Kwiatkowska},\ and\ \citenamefont
  {Hamant}}]{Uyttewaal2012}%
  \BibitemOpen
  \bibfield  {author} {\bibinfo {author} {\bibfnamefont {M.}~\bibnamefont
  {Uyttewaal}}, \bibinfo {author} {\bibfnamefont {A.}~\bibnamefont {Burian}},
  \bibinfo {author} {\bibfnamefont {K.}~\bibnamefont {Alim}}, \bibinfo {author}
  {\bibfnamefont {B.}~\bibnamefont {Landrein}}, \bibinfo {author}
  {\bibfnamefont {D.}~\bibnamefont {Borowska-Wykret}}, \bibinfo {author}
  {\bibfnamefont {A.}~\bibnamefont {Dedieu}}, \bibinfo {author} {\bibfnamefont
  {A.}~\bibnamefont {Peaucelle}}, \bibinfo {author} {\bibfnamefont
  {M.}~\bibnamefont {Ludynia}}, \bibinfo {author} {\bibfnamefont
  {J.}~\bibnamefont {Traas}}, \bibinfo {author} {\bibfnamefont
  {A.}~\bibnamefont {Boudaoud}}, \bibinfo {author} {\bibfnamefont
  {D.}~\bibnamefont {Kwiatkowska}}, \ and\ \bibinfo {author} {\bibfnamefont
  {O.}~\bibnamefont {Hamant}},\ }\href@noop {} {\bibfield  {journal} {\bibinfo
  {journal} {Cell}\ }\textbf {\bibinfo {volume} {149}},\ \bibinfo {pages} {439
  } (\bibinfo {year} {2012})}\BibitemShut {NoStop}%
\bibitem [{\citenamefont {Streichan}\ \emph {et~al.}(2014)\citenamefont
  {Streichan}, \citenamefont {Hoerner}, \citenamefont {Schneidt}, \citenamefont
  {Holzer},\ and\ \citenamefont {Hufnagel}}]{Streichan2014}%
  \BibitemOpen
  \bibfield  {author} {\bibinfo {author} {\bibfnamefont {S.~J.}\ \bibnamefont
  {Streichan}}, \bibinfo {author} {\bibfnamefont {C.~R.}\ \bibnamefont
  {Hoerner}}, \bibinfo {author} {\bibfnamefont {T.}~\bibnamefont {Schneidt}},
  \bibinfo {author} {\bibfnamefont {D.}~\bibnamefont {Holzer}}, \ and\ \bibinfo
  {author} {\bibfnamefont {L.}~\bibnamefont {Hufnagel}},\ }\href@noop {}
  {\bibfield  {journal} {\bibinfo  {journal} {Proc. Natl. Acad. Sci. U.S.A.}\
  }\textbf {\bibinfo {volume} {111}},\ \bibinfo {pages} {5586} (\bibinfo {year}
  {2014})}\BibitemShut {NoStop}%
\bibitem [{\citenamefont {LeGoff}\ and\ \citenamefont
  {Lecuit}(2015)}]{LeGoff2015}%
  \BibitemOpen
  \bibfield  {author} {\bibinfo {author} {\bibfnamefont {L.}~\bibnamefont
  {LeGoff}}\ and\ \bibinfo {author} {\bibfnamefont {T.}~\bibnamefont
  {Lecuit}},\ }\href@noop {} {\bibfield  {journal} {\bibinfo  {journal} {Cold
  Spring Harbor Perspect. Biol.}\ }\textbf {\bibinfo {volume} {8}},\ \bibinfo
  {pages} {a019232} (\bibinfo {year} {2015})}\BibitemShut {NoStop}%
\bibitem [{\citenamefont {Eder}\ \emph {et~al.}(2017)\citenamefont {Eder},
  \citenamefont {Aegerter},\ and\ \citenamefont {Basler}}]{Eder2017}%
  \BibitemOpen
  \bibfield  {author} {\bibinfo {author} {\bibfnamefont {D.}~\bibnamefont
  {Eder}}, \bibinfo {author} {\bibfnamefont {C.}~\bibnamefont {Aegerter}}, \
  and\ \bibinfo {author} {\bibfnamefont {K.}~\bibnamefont {Basler}},\
  }\href@noop {} {\bibfield  {journal} {\bibinfo  {journal} {Mech. Dev.}\
  }\textbf {\bibinfo {volume} {144}},\ \bibinfo {pages} {53 } (\bibinfo {year}
  {2017})}\BibitemShut {NoStop}%
\bibitem [{\citenamefont {Helmlinger}\ \emph {et~al.}(1997)\citenamefont
  {Helmlinger}, \citenamefont {Netti}, \citenamefont {Lichtenbeld},
  \citenamefont {Melder},\ and\ \citenamefont {Jain}}]{Helmlinger1997}%
  \BibitemOpen
  \bibfield  {author} {\bibinfo {author} {\bibfnamefont {G.}~\bibnamefont
  {Helmlinger}}, \bibinfo {author} {\bibfnamefont {P.~A.}\ \bibnamefont
  {Netti}}, \bibinfo {author} {\bibfnamefont {H.~C.}\ \bibnamefont
  {Lichtenbeld}}, \bibinfo {author} {\bibfnamefont {R.~J.}\ \bibnamefont
  {Melder}}, \ and\ \bibinfo {author} {\bibfnamefont {R.~K.}\ \bibnamefont
  {Jain}},\ }\href@noop {} {\bibfield  {journal} {\bibinfo  {journal} {Nat.
  Biotechnol.}\ }\textbf {\bibinfo {volume} {15}},\ \bibinfo {pages} {778}
  (\bibinfo {year} {1997})}\BibitemShut {NoStop}%
\bibitem [{\citenamefont {Gordon}\ \emph {et~al.}(2003)\citenamefont {Gordon},
  \citenamefont {Valentine}, \citenamefont {Gardel}, \citenamefont
  {Andor-Ardo}, \citenamefont {Dennison}, \citenamefont {Bogdanov},
  \citenamefont {Weitz},\ and\ \citenamefont {Deisboeck}}]{Gordon2003}%
  \BibitemOpen
  \bibfield  {author} {\bibinfo {author} {\bibfnamefont {V.~D.}\ \bibnamefont
  {Gordon}}, \bibinfo {author} {\bibfnamefont {M.~T.}\ \bibnamefont
  {Valentine}}, \bibinfo {author} {\bibfnamefont {M.~L.}\ \bibnamefont
  {Gardel}}, \bibinfo {author} {\bibfnamefont {D.}~\bibnamefont {Andor-Ardo}},
  \bibinfo {author} {\bibfnamefont {S.}~\bibnamefont {Dennison}}, \bibinfo
  {author} {\bibfnamefont {A.~A.}\ \bibnamefont {Bogdanov}}, \bibinfo {author}
  {\bibfnamefont {D.~A.}\ \bibnamefont {Weitz}}, \ and\ \bibinfo {author}
  {\bibfnamefont {T.~S.}\ \bibnamefont {Deisboeck}},\ }\href@noop {} {\bibfield
   {journal} {\bibinfo  {journal} {Exp. Cell Res.}\ }\textbf {\bibinfo {volume}
  {289}},\ \bibinfo {pages} {58} (\bibinfo {year} {2003})}\BibitemShut
  {NoStop}%
\bibitem [{\citenamefont {Kaufman}\ \emph {et~al.}(2005)\citenamefont
  {Kaufman}, \citenamefont {Brangwynne}, \citenamefont {Kasza}, \citenamefont
  {Filippidi}, \citenamefont {Gordon}, \citenamefont {Deisboeck},\ and\
  \citenamefont {Weitz}}]{Kaufman2005}%
  \BibitemOpen
  \bibfield  {author} {\bibinfo {author} {\bibfnamefont {L.~J.}\ \bibnamefont
  {Kaufman}}, \bibinfo {author} {\bibfnamefont {C.~P.}\ \bibnamefont
  {Brangwynne}}, \bibinfo {author} {\bibfnamefont {K.~E.}\ \bibnamefont
  {Kasza}}, \bibinfo {author} {\bibfnamefont {E.}~\bibnamefont {Filippidi}},
  \bibinfo {author} {\bibfnamefont {V.~D.}\ \bibnamefont {Gordon}}, \bibinfo
  {author} {\bibfnamefont {T.~S.}\ \bibnamefont {Deisboeck}}, \ and\ \bibinfo
  {author} {\bibfnamefont {D.~A.}\ \bibnamefont {Weitz}},\ }\href@noop {}
  {\bibfield  {journal} {\bibinfo  {journal} {Biophys. J.}\ }\textbf {\bibinfo
  {volume} {89}},\ \bibinfo {pages} {635} (\bibinfo {year} {2005})}\BibitemShut
  {NoStop}%
\bibitem [{\citenamefont {Alessandri}\ \emph {et~al.}(2013)\citenamefont
  {Alessandri}, \citenamefont {Sarangi}, \citenamefont {Gurchenkov},
  \citenamefont {Sinha}, \citenamefont {Kiesling}, \citenamefont {Fetler},
  \citenamefont {Rico}, \citenamefont {Scheuring}, \citenamefont {Lamaze},
  \citenamefont {Simon}, \citenamefont {Geraldo}, \citenamefont {Vignjevic},
  \citenamefont {Domejean}, \citenamefont {Rolland}, \citenamefont {Funfak},
  \citenamefont {Bibette}, \citenamefont {Bremond},\ and\ \citenamefont
  {Nassoy}}]{Alessandri2013}%
  \BibitemOpen
  \bibfield  {author} {\bibinfo {author} {\bibfnamefont {K.}~\bibnamefont
  {Alessandri}}, \bibinfo {author} {\bibfnamefont {B.~R.}\ \bibnamefont
  {Sarangi}}, \bibinfo {author} {\bibfnamefont {V.~V.}\ \bibnamefont
  {Gurchenkov}}, \bibinfo {author} {\bibfnamefont {B.}~\bibnamefont {Sinha}},
  \bibinfo {author} {\bibfnamefont {T.~R.}\ \bibnamefont {Kiesling}}, \bibinfo
  {author} {\bibfnamefont {L.}~\bibnamefont {Fetler}}, \bibinfo {author}
  {\bibfnamefont {F.}~\bibnamefont {Rico}}, \bibinfo {author} {\bibfnamefont
  {S.}~\bibnamefont {Scheuring}}, \bibinfo {author} {\bibfnamefont
  {C.}~\bibnamefont {Lamaze}}, \bibinfo {author} {\bibfnamefont
  {A.}~\bibnamefont {Simon}}, \bibinfo {author} {\bibfnamefont
  {S.}~\bibnamefont {Geraldo}}, \bibinfo {author} {\bibfnamefont
  {D.}~\bibnamefont {Vignjevic}}, \bibinfo {author} {\bibfnamefont
  {H.}~\bibnamefont {Domejean}}, \bibinfo {author} {\bibfnamefont
  {L.}~\bibnamefont {Rolland}}, \bibinfo {author} {\bibfnamefont
  {A.}~\bibnamefont {Funfak}}, \bibinfo {author} {\bibfnamefont
  {J.}~\bibnamefont {Bibette}}, \bibinfo {author} {\bibfnamefont
  {N.}~\bibnamefont {Bremond}}, \ and\ \bibinfo {author} {\bibfnamefont
  {P.}~\bibnamefont {Nassoy}},\ }\href@noop {} {\bibfield  {journal} {\bibinfo
  {journal} {Proc. Natl. Acad. Sci. U.S.A.}\ }\textbf {\bibinfo {volume}
  {110}},\ \bibinfo {pages} {14843} (\bibinfo {year} {2013})}\BibitemShut
  {NoStop}%
\bibitem [{\citenamefont {Domejean}\ \emph {et~al.}(2017)\citenamefont
  {Domejean}, \citenamefont {Saint~Pierre}, \citenamefont {Funfak},
  \citenamefont {Atrux-Tallau}, \citenamefont {Alessandri}, \citenamefont
  {Nassoy}, \citenamefont {Bibette},\ and\ \citenamefont
  {Bremond}}]{Domejean2017}%
  \BibitemOpen
  \bibfield  {author} {\bibinfo {author} {\bibfnamefont {H.}~\bibnamefont
  {Domejean}}, \bibinfo {author} {\bibfnamefont {M.~d. l.~M.}\ \bibnamefont
  {Saint~Pierre}}, \bibinfo {author} {\bibfnamefont {A.}~\bibnamefont
  {Funfak}}, \bibinfo {author} {\bibfnamefont {N.}~\bibnamefont
  {Atrux-Tallau}}, \bibinfo {author} {\bibfnamefont {K.}~\bibnamefont
  {Alessandri}}, \bibinfo {author} {\bibfnamefont {P.}~\bibnamefont {Nassoy}},
  \bibinfo {author} {\bibfnamefont {J.}~\bibnamefont {Bibette}}, \ and\
  \bibinfo {author} {\bibfnamefont {N.}~\bibnamefont {Bremond}},\ }\href@noop
  {} {\bibfield  {journal} {\bibinfo  {journal} {Lab. Chip}\ }\textbf {\bibinfo
  {volume} {{17}}},\ \bibinfo {pages} {{110}} (\bibinfo {year}
  {{2017}})}\BibitemShut {NoStop}%
\bibitem [{\citenamefont {Montel}\ \emph {et~al.}(2011)\citenamefont {Montel},
  \citenamefont {Delarue}, \citenamefont {Elgeti}, \citenamefont {Malaquin},
  \citenamefont {Basan}, \citenamefont {Risler}, \citenamefont {Cabane},
  \citenamefont {Vignjevic}, \citenamefont {Prost}, \citenamefont {Cappello},\
  and\ \citenamefont {Joanny}}]{Montel2011}%
  \BibitemOpen
  \bibfield  {author} {\bibinfo {author} {\bibfnamefont {F.}~\bibnamefont
  {Montel}}, \bibinfo {author} {\bibfnamefont {M.}~\bibnamefont {Delarue}},
  \bibinfo {author} {\bibfnamefont {J.}~\bibnamefont {Elgeti}}, \bibinfo
  {author} {\bibfnamefont {L.}~\bibnamefont {Malaquin}}, \bibinfo {author}
  {\bibfnamefont {M.}~\bibnamefont {Basan}}, \bibinfo {author} {\bibfnamefont
  {T.}~\bibnamefont {Risler}}, \bibinfo {author} {\bibfnamefont
  {B.}~\bibnamefont {Cabane}}, \bibinfo {author} {\bibfnamefont
  {D.}~\bibnamefont {Vignjevic}}, \bibinfo {author} {\bibfnamefont
  {J.}~\bibnamefont {Prost}}, \bibinfo {author} {\bibfnamefont
  {G.}~\bibnamefont {Cappello}}, \ and\ \bibinfo {author} {\bibfnamefont
  {J.~F.}\ \bibnamefont {Joanny}},\ }\href@noop {} {\bibfield  {journal}
  {\bibinfo  {journal} {Phys. Rev. Lett.}\ }\textbf {\bibinfo {volume} {107}},\
  \bibinfo {pages} {188102} (\bibinfo {year} {2011})}\BibitemShut {NoStop}%
\bibitem [{\citenamefont {Montel}\ \emph {et~al.}(2012)\citenamefont {Montel},
  \citenamefont {Delarue}, \citenamefont {Elgeti}, \citenamefont {Vignjevic},
  \citenamefont {Cappello},\ and\ \citenamefont {Prost}}]{Montel2012}%
  \BibitemOpen
  \bibfield  {author} {\bibinfo {author} {\bibfnamefont {F.}~\bibnamefont
  {Montel}}, \bibinfo {author} {\bibfnamefont {M.}~\bibnamefont {Delarue}},
  \bibinfo {author} {\bibfnamefont {J.}~\bibnamefont {Elgeti}}, \bibinfo
  {author} {\bibfnamefont {D.}~\bibnamefont {Vignjevic}}, \bibinfo {author}
  {\bibfnamefont {G.}~\bibnamefont {Cappello}}, \ and\ \bibinfo {author}
  {\bibfnamefont {J.}~\bibnamefont {Prost}},\ }\href@noop {} {\bibfield
  {journal} {\bibinfo  {journal} {New J. Phys.}\ }\textbf {\bibinfo {volume}
  {14}},\ \bibinfo {pages} {055008} (\bibinfo {year} {2012})}\BibitemShut
  {NoStop}%
\bibitem [{\citenamefont {Delarue}\ \emph {et~al.}(2013)\citenamefont
  {Delarue}, \citenamefont {Montel}, \citenamefont {Caen}, \citenamefont
  {Elgeti}, \citenamefont {Siaugue}, \citenamefont {Vignjevic}, \citenamefont
  {Prost}, \citenamefont {Joanny},\ and\ \citenamefont
  {Cappello}}]{Delarue2013}%
  \BibitemOpen
  \bibfield  {author} {\bibinfo {author} {\bibfnamefont {M.}~\bibnamefont
  {Delarue}}, \bibinfo {author} {\bibfnamefont {F.}~\bibnamefont {Montel}},
  \bibinfo {author} {\bibfnamefont {O.}~\bibnamefont {Caen}}, \bibinfo {author}
  {\bibfnamefont {J.}~\bibnamefont {Elgeti}}, \bibinfo {author} {\bibfnamefont
  {J.-M.}\ \bibnamefont {Siaugue}}, \bibinfo {author} {\bibfnamefont
  {D.}~\bibnamefont {Vignjevic}}, \bibinfo {author} {\bibfnamefont
  {J.}~\bibnamefont {Prost}}, \bibinfo {author} {\bibfnamefont {J.-F.}\
  \bibnamefont {Joanny}}, \ and\ \bibinfo {author} {\bibfnamefont
  {G.}~\bibnamefont {Cappello}},\ }\href@noop {} {\bibfield  {journal}
  {\bibinfo  {journal} {Phys. Rev. Lett.}\ }\textbf {\bibinfo {volume} {110}},\
  \bibinfo {pages} {138103} (\bibinfo {year} {2013})}\BibitemShut {NoStop}%
\bibitem [{\citenamefont {Taloni}\ \emph {et~al.}(2014)\citenamefont {Taloni},
  \citenamefont {Alemi}, \citenamefont {Ciusani}, \citenamefont {Sethna},
  \citenamefont {Zapperi},\ and\ \citenamefont {{La Porta}}}]{Taloni2014}%
  \BibitemOpen
  \bibfield  {author} {\bibinfo {author} {\bibfnamefont {A.}~\bibnamefont
  {Taloni}}, \bibinfo {author} {\bibfnamefont {A.~A.}\ \bibnamefont {Alemi}},
  \bibinfo {author} {\bibfnamefont {E.}~\bibnamefont {Ciusani}}, \bibinfo
  {author} {\bibfnamefont {J.~P.}\ \bibnamefont {Sethna}}, \bibinfo {author}
  {\bibfnamefont {S.}~\bibnamefont {Zapperi}}, \ and\ \bibinfo {author}
  {\bibfnamefont {C.~A.~M.}\ \bibnamefont {{La Porta}}},\ }\href@noop {}
  {\bibfield  {journal} {\bibinfo  {journal} {PLoS One}\ }\textbf {\bibinfo
  {volume} {9}},\ \bibinfo {pages} {e94229} (\bibinfo {year}
  {2014})}\BibitemShut {NoStop}%
\bibitem [{\citenamefont {Morata}\ and\ \citenamefont
  {Ripoll}(1975)}]{Morata1975}%
  \BibitemOpen
  \bibfield  {author} {\bibinfo {author} {\bibfnamefont {G.}~\bibnamefont
  {Morata}}\ and\ \bibinfo {author} {\bibfnamefont {P.}~\bibnamefont
  {Ripoll}},\ }\href@noop {} {\bibfield  {journal} {\bibinfo  {journal} {Dev.
  Biol.}\ }\textbf {\bibinfo {volume} {42}},\ \bibinfo {pages} {211 } (\bibinfo
  {year} {1975})}\BibitemShut {NoStop}%
\bibitem [{\citenamefont {Diaz}\ and\ \citenamefont {Moreno}(2005)}]{Diaz2005}%
  \BibitemOpen
  \bibfield  {author} {\bibinfo {author} {\bibfnamefont {B.}~\bibnamefont
  {Diaz}}\ and\ \bibinfo {author} {\bibfnamefont {E.}~\bibnamefont {Moreno}},\
  }\href@noop {} {\bibfield  {journal} {\bibinfo  {journal} {Exp. Cell. Res.}\
  }\textbf {\bibinfo {volume} {306}},\ \bibinfo {pages} {317 } (\bibinfo {year}
  {2005})}\BibitemShut {NoStop}%
\bibitem [{\citenamefont {Moolgavkar}\ and\ \citenamefont
  {Luebeck}(2003)}]{Moolgavkar2003}%
  \BibitemOpen
  \bibfield  {author} {\bibinfo {author} {\bibfnamefont {S.~H.}\ \bibnamefont
  {Moolgavkar}}\ and\ \bibinfo {author} {\bibfnamefont {E.~G.}\ \bibnamefont
  {Luebeck}},\ }\href@noop {} {\bibfield  {journal} {\bibinfo  {journal}
  {Genes. Chromosomes Cancer}\ }\textbf {\bibinfo {volume} {38}},\ \bibinfo
  {pages} {302} (\bibinfo {year} {2003})}\BibitemShut {NoStop}%
\bibitem [{\citenamefont {Meza}\ and\ \citenamefont {Chang}(2015)}]{Meza2015}%
  \BibitemOpen
  \bibfield  {author} {\bibinfo {author} {\bibfnamefont {R.}~\bibnamefont
  {Meza}}\ and\ \bibinfo {author} {\bibfnamefont {J.~T.}\ \bibnamefont
  {Chang}},\ }\href@noop {} {\bibfield  {journal} {\bibinfo  {journal} {BMC
  Public Health}\ }\textbf {\bibinfo {volume} {15}},\ \bibinfo {pages} {789}
  (\bibinfo {year} {2015})}\BibitemShut {NoStop}%
\bibitem [{\citenamefont {Hufnagel}\ \emph {et~al.}(2007)\citenamefont
  {Hufnagel}, \citenamefont {Teleman}, \citenamefont {Rouault}, \citenamefont
  {Cohen},\ and\ \citenamefont {Shraiman}}]{Hufnagel2007}%
  \BibitemOpen
  \bibfield  {author} {\bibinfo {author} {\bibfnamefont {L.}~\bibnamefont
  {Hufnagel}}, \bibinfo {author} {\bibfnamefont {A.~A.}\ \bibnamefont
  {Teleman}}, \bibinfo {author} {\bibfnamefont {H.}~\bibnamefont {Rouault}},
  \bibinfo {author} {\bibfnamefont {S.~M.}\ \bibnamefont {Cohen}}, \ and\
  \bibinfo {author} {\bibfnamefont {B.~I.}\ \bibnamefont {Shraiman}},\
  }\href@noop {} {\bibfield  {journal} {\bibinfo  {journal} {Proc. Natl. Acad.
  Sci. U. S. A.}\ }\textbf {\bibinfo {volume} {104}},\ \bibinfo {pages} {3835}
  (\bibinfo {year} {2007})}\BibitemShut {NoStop}%
\bibitem [{\citenamefont {Basan}\ \emph {et~al.}(2009)\citenamefont {Basan},
  \citenamefont {Risler}, \citenamefont {Joanny}, \citenamefont
  {Sastre-Garau},\ and\ \citenamefont {Prost}}]{Basan2009}%
  \BibitemOpen
  \bibfield  {author} {\bibinfo {author} {\bibfnamefont {M.}~\bibnamefont
  {Basan}}, \bibinfo {author} {\bibfnamefont {T.}~\bibnamefont {Risler}},
  \bibinfo {author} {\bibfnamefont {J.-F.}\ \bibnamefont {Joanny}}, \bibinfo
  {author} {\bibfnamefont {X.}~\bibnamefont {Sastre-Garau}}, \ and\ \bibinfo
  {author} {\bibfnamefont {J.}~\bibnamefont {Prost}},\ }\href@noop {}
  {\bibfield  {journal} {\bibinfo  {journal} {HFSP J}\ }\textbf {\bibinfo
  {volume} {3}},\ \bibinfo {pages} {265} (\bibinfo {year} {2009})}\BibitemShut
  {NoStop}%
\bibitem [{\citenamefont {Basan}\ \emph {et~al.}(2011)\citenamefont {Basan},
  \citenamefont {Prost}, \citenamefont {Joanny},\ and\ \citenamefont
  {Elgeti}}]{Basan2011a}%
  \BibitemOpen
  \bibfield  {author} {\bibinfo {author} {\bibfnamefont {M.}~\bibnamefont
  {Basan}}, \bibinfo {author} {\bibfnamefont {J.}~\bibnamefont {Prost}},
  \bibinfo {author} {\bibfnamefont {J.-F.}\ \bibnamefont {Joanny}}, \ and\
  \bibinfo {author} {\bibfnamefont {J.}~\bibnamefont {Elgeti}},\ }\href@noop {}
  {\bibfield  {journal} {\bibinfo  {journal} {Phys. Biol.}\ }\textbf {\bibinfo
  {volume} {8}},\ \bibinfo {pages} {026014} (\bibinfo {year}
  {2011})}\BibitemShut {NoStop}%
\bibitem [{\citenamefont {Podewitz}\ \emph {et~al.}(2015)\citenamefont
  {Podewitz}, \citenamefont {Delarue},\ and\ \citenamefont
  {Elgeti}}]{Podewitz2015}%
  \BibitemOpen
  \bibfield  {author} {\bibinfo {author} {\bibfnamefont {N.}~\bibnamefont
  {Podewitz}}, \bibinfo {author} {\bibfnamefont {M.}~\bibnamefont {Delarue}}, \
  and\ \bibinfo {author} {\bibfnamefont {J.}~\bibnamefont {Elgeti}},\
  }\href@noop {} {\bibfield  {journal} {\bibinfo  {journal} {Europhys. Lett.}\
  }\textbf {\bibinfo {volume} {109}},\ \bibinfo {pages} {58005} (\bibinfo
  {year} {2015})}\BibitemShut {NoStop}%
\bibitem [{\citenamefont {Podewitz}\ \emph {et~al.}(2016)\citenamefont
  {Podewitz}, \citenamefont {J{\"u}licher}, \citenamefont {Gompper},\ and\
  \citenamefont {Elgeti}}]{Podewitz2016}%
  \BibitemOpen
  \bibfield  {author} {\bibinfo {author} {\bibfnamefont {N.}~\bibnamefont
  {Podewitz}}, \bibinfo {author} {\bibfnamefont {F.}~\bibnamefont
  {J{\"u}licher}}, \bibinfo {author} {\bibfnamefont {G.}~\bibnamefont
  {Gompper}}, \ and\ \bibinfo {author} {\bibfnamefont {J.}~\bibnamefont
  {Elgeti}},\ }\href@noop {} {\bibfield  {journal} {\bibinfo  {journal} {New J.
  Phys.}\ }\textbf {\bibinfo {volume} {18}},\ \bibinfo {pages} {083020}
  (\bibinfo {year} {2016})}\BibitemShut {NoStop}%
\bibitem [{\citenamefont {Jagiella}\ \emph {et~al.}(2016)\citenamefont
  {Jagiella}, \citenamefont {Mueller}, \citenamefont {Mueller}, \citenamefont
  {Vignon-Clementel},\ and\ \citenamefont {Drasdo}}]{Jagiella2016}%
  \BibitemOpen
  \bibfield  {author} {\bibinfo {author} {\bibfnamefont {N.}~\bibnamefont
  {Jagiella}}, \bibinfo {author} {\bibfnamefont {B.}~\bibnamefont {Mueller}},
  \bibinfo {author} {\bibfnamefont {M.}~\bibnamefont {Mueller}}, \bibinfo
  {author} {\bibfnamefont {I.~E.}\ \bibnamefont {Vignon-Clementel}}, \ and\
  \bibinfo {author} {\bibfnamefont {D.}~\bibnamefont {Drasdo}},\ }\href@noop {}
  {\bibfield  {journal} {\bibinfo  {journal} {PLoS Comput. Biol.}\ }\textbf
  {\bibinfo {volume} {12}},\ \bibinfo {pages} {1} (\bibinfo {year}
  {2016})}\BibitemShut {NoStop}%
\bibitem [{\citenamefont {Van~Liedekerke}\ \emph {et~al.}(2015)\citenamefont
  {Van~Liedekerke}, \citenamefont {Palm}, \citenamefont {Jagiella},\ and\
  \citenamefont {Drasdo}}]{VanLiedekerke2015}%
  \BibitemOpen
  \bibfield  {author} {\bibinfo {author} {\bibfnamefont {P.}~\bibnamefont
  {Van~Liedekerke}}, \bibinfo {author} {\bibfnamefont {M.~M.}\ \bibnamefont
  {Palm}}, \bibinfo {author} {\bibfnamefont {N.}~\bibnamefont {Jagiella}}, \
  and\ \bibinfo {author} {\bibfnamefont {D.}~\bibnamefont {Drasdo}},\
  }\href@noop {} {\bibfield  {journal} {\bibinfo  {journal} {Comput. Part.
  Mech.}\ }\textbf {\bibinfo {volume} {2}},\ \bibinfo {pages} {401} (\bibinfo
  {year} {2015})}\BibitemShut {NoStop}%
\bibitem [{\citenamefont {Kobayashi}\ \emph {et~al.}(2018)\citenamefont
  {Kobayashi}, \citenamefont {Yasugahira}, \citenamefont {Kitahata},
  \citenamefont {Watanabe}, \citenamefont {Natsuga},\ and\ \citenamefont
  {Nagayama}}]{Kobayashi2018}%
  \BibitemOpen
  \bibfield  {author} {\bibinfo {author} {\bibfnamefont {Y.}~\bibnamefont
  {Kobayashi}}, \bibinfo {author} {\bibfnamefont {Y.}~\bibnamefont
  {Yasugahira}}, \bibinfo {author} {\bibfnamefont {H.}~\bibnamefont
  {Kitahata}}, \bibinfo {author} {\bibfnamefont {M.}~\bibnamefont {Watanabe}},
  \bibinfo {author} {\bibfnamefont {K.}~\bibnamefont {Natsuga}}, \ and\
  \bibinfo {author} {\bibfnamefont {M.}~\bibnamefont {Nagayama}},\ }\href@noop
  {} {\bibfield  {journal} {\bibinfo  {journal} {npj Comput. Mater.}\ }\textbf
  {\bibinfo {volume} {4}},\ \bibinfo {pages} {45} (\bibinfo {year}
  {2018})}\BibitemShut {NoStop}%
\bibitem [{\citenamefont {Ranft}\ \emph {et~al.}(2010)\citenamefont {Ranft},
  \citenamefont {Basan}, \citenamefont {Elgeti}, \citenamefont {Joanny},
  \citenamefont {Prost},\ and\ \citenamefont {J{\"u}licher}}]{Ranft2010}%
  \BibitemOpen
  \bibfield  {author} {\bibinfo {author} {\bibfnamefont {J.}~\bibnamefont
  {Ranft}}, \bibinfo {author} {\bibfnamefont {M.}~\bibnamefont {Basan}},
  \bibinfo {author} {\bibfnamefont {J.}~\bibnamefont {Elgeti}}, \bibinfo
  {author} {\bibfnamefont {J.-F.}\ \bibnamefont {Joanny}}, \bibinfo {author}
  {\bibfnamefont {J.}~\bibnamefont {Prost}}, \ and\ \bibinfo {author}
  {\bibfnamefont {F.}~\bibnamefont {J{\"u}licher}},\ }\href@noop {} {\bibfield
  {journal} {\bibinfo  {journal} {Proc. Natl. Acad. Sci. U. S. A.}\ }\textbf
  {\bibinfo {volume} {107}},\ \bibinfo {pages} {20863} (\bibinfo {year}
  {2010})}\BibitemShut {NoStop}%
\bibitem [{\citenamefont {Khalilgharibi}\ \emph {et~al.}(2016)\citenamefont
  {Khalilgharibi}, \citenamefont {Fouchard}, \citenamefont {Recho},
  \citenamefont {Charras},\ and\ \citenamefont {Kabla}}]{Khalilgharibi2016}%
  \BibitemOpen
  \bibfield  {author} {\bibinfo {author} {\bibfnamefont {N.}~\bibnamefont
  {Khalilgharibi}}, \bibinfo {author} {\bibfnamefont {J.}~\bibnamefont
  {Fouchard}}, \bibinfo {author} {\bibfnamefont {P.}~\bibnamefont {Recho}},
  \bibinfo {author} {\bibfnamefont {G.}~\bibnamefont {Charras}}, \ and\
  \bibinfo {author} {\bibfnamefont {A.}~\bibnamefont {Kabla}},\ }\href@noop {}
  {\bibfield  {journal} {\bibinfo  {journal} {Curr. Opin. Cell Biol.}\ }\textbf
  {\bibinfo {volume} {{42}}},\ \bibinfo {pages} {{113}} (\bibinfo {year}
  {{2016}})}\BibitemShut {NoStop}%
\bibitem [{\citenamefont {Matoz-Fernandez}\ \emph {et~al.}(2017)\citenamefont
  {Matoz-Fernandez}, \citenamefont {Agoritsas}, \citenamefont {Barrat},
  \citenamefont {Bertin},\ and\ \citenamefont
  {Martens}}]{Matoz-Fernandez2017a}%
  \BibitemOpen
  \bibfield  {author} {\bibinfo {author} {\bibfnamefont {D.~A.}\ \bibnamefont
  {Matoz-Fernandez}}, \bibinfo {author} {\bibfnamefont {E.}~\bibnamefont
  {Agoritsas}}, \bibinfo {author} {\bibfnamefont {J.-L.}\ \bibnamefont
  {Barrat}}, \bibinfo {author} {\bibfnamefont {E.}~\bibnamefont {Bertin}}, \
  and\ \bibinfo {author} {\bibfnamefont {K.}~\bibnamefont {Martens}},\
  }\href@noop {} {\bibfield  {journal} {\bibinfo  {journal} {Phys. Rev. Lett.}\
  }\textbf {\bibinfo {volume} {{118}}} (\bibinfo {year} {{2017}})}\BibitemShut
  {NoStop}%
\bibitem [{\citenamefont {Phillips}\ and\ \citenamefont
  {Steinberg}(1978)}]{Phillips1978}%
  \BibitemOpen
  \bibfield  {author} {\bibinfo {author} {\bibfnamefont {H.}~\bibnamefont
  {Phillips}}\ and\ \bibinfo {author} {\bibfnamefont {M.}~\bibnamefont
  {Steinberg}},\ }\href@noop {} {\bibfield  {journal} {\bibinfo  {journal} {J.
  Cell Sci.}\ }\textbf {\bibinfo {volume} {30}},\ \bibinfo {pages} {1}
  (\bibinfo {year} {1978})}\BibitemShut {NoStop}%
\bibitem [{\citenamefont {Guevorkian}\ \emph {et~al.}(2010)\citenamefont
  {Guevorkian}, \citenamefont {Colbert}, \citenamefont {Durth}, \citenamefont
  {Dufour},\ and\ \citenamefont {Brochard-Wyart}}]{Guevorkian2010}%
  \BibitemOpen
  \bibfield  {author} {\bibinfo {author} {\bibfnamefont {K.}~\bibnamefont
  {Guevorkian}}, \bibinfo {author} {\bibfnamefont {M.-J.}\ \bibnamefont
  {Colbert}}, \bibinfo {author} {\bibfnamefont {M.}~\bibnamefont {Durth}},
  \bibinfo {author} {\bibfnamefont {S.}~\bibnamefont {Dufour}}, \ and\ \bibinfo
  {author} {\bibfnamefont {F.}~\bibnamefont {Brochard-Wyart}},\ }\href@noop {}
  {\bibfield  {journal} {\bibinfo  {journal} {Phys. Rev. Lett.}\ }\textbf
  {\bibinfo {volume} {104}},\ \bibinfo {pages} {218101} (\bibinfo {year}
  {2010})}\BibitemShut {NoStop}%
\bibitem [{\citenamefont {Gonzalez-Rodriguez}\ \emph
  {et~al.}(2013)\citenamefont {Gonzalez-Rodriguez}, \citenamefont {Bonnemay},
  \citenamefont {Elgeti}, \citenamefont {Dufour}, \citenamefont {Cuvelier},\
  and\ \citenamefont {Brochard-Wyart}}]{Gonzalez-Rodriguez2013}%
  \BibitemOpen
  \bibfield  {author} {\bibinfo {author} {\bibfnamefont {D.}~\bibnamefont
  {Gonzalez-Rodriguez}}, \bibinfo {author} {\bibfnamefont {L.}~\bibnamefont
  {Bonnemay}}, \bibinfo {author} {\bibfnamefont {J.}~\bibnamefont {Elgeti}},
  \bibinfo {author} {\bibfnamefont {S.}~\bibnamefont {Dufour}}, \bibinfo
  {author} {\bibfnamefont {D.}~\bibnamefont {Cuvelier}}, \ and\ \bibinfo
  {author} {\bibfnamefont {F.}~\bibnamefont {Brochard-Wyart}},\ }\href@noop {}
  {\bibfield  {journal} {\bibinfo  {journal} {Soft Matter}\ }\textbf {\bibinfo
  {volume} {9}},\ \bibinfo {pages} {2282} (\bibinfo {year} {2013})}\BibitemShut
  {NoStop}%
\bibitem [{\citenamefont {LeGoff}\ \emph {et~al.}(2013)\citenamefont {LeGoff},
  \citenamefont {Rouault},\ and\ \citenamefont {Lecuit}}]{LeGoff2013}%
  \BibitemOpen
  \bibfield  {author} {\bibinfo {author} {\bibfnamefont {L.}~\bibnamefont
  {LeGoff}}, \bibinfo {author} {\bibfnamefont {H.}~\bibnamefont {Rouault}}, \
  and\ \bibinfo {author} {\bibfnamefont {T.}~\bibnamefont {Lecuit}},\
  }\href@noop {} {\bibfield  {journal} {\bibinfo  {journal} {Development}\
  }\textbf {\bibinfo {volume} {140}},\ \bibinfo {pages} {4051} (\bibinfo {year}
  {2013})}\BibitemShut {NoStop}%
\bibitem [{\citenamefont {Mao}\ \emph {et~al.}(2013)\citenamefont {Mao},
  \citenamefont {Tournier}, \citenamefont {Hoppe}, \citenamefont {Kester},
  \citenamefont {Thompson},\ and\ \citenamefont {Tapon}}]{Mao2013}%
  \BibitemOpen
  \bibfield  {author} {\bibinfo {author} {\bibfnamefont {Y.}~\bibnamefont
  {Mao}}, \bibinfo {author} {\bibfnamefont {A.~L.}\ \bibnamefont {Tournier}},
  \bibinfo {author} {\bibfnamefont {A.}~\bibnamefont {Hoppe}}, \bibinfo
  {author} {\bibfnamefont {L.}~\bibnamefont {Kester}}, \bibinfo {author}
  {\bibfnamefont {B.~J.}\ \bibnamefont {Thompson}}, \ and\ \bibinfo {author}
  {\bibfnamefont {N.}~\bibnamefont {Tapon}},\ }\href@noop {} {\bibfield
  {journal} {\bibinfo  {journal} {The EMBO journal}\ }\textbf {\bibinfo
  {volume} {32}},\ \bibinfo {pages} {2790} (\bibinfo {year}
  {2013})}\BibitemShut {NoStop}%
\bibitem [{\citenamefont {Pan}\ \emph {et~al.}(2016)\citenamefont {Pan},
  \citenamefont {Heemskerk}, \citenamefont {Ibar}, \citenamefont {Shraiman},\
  and\ \citenamefont {Irvine}}]{Pan2016}%
  \BibitemOpen
  \bibfield  {author} {\bibinfo {author} {\bibfnamefont {Y.}~\bibnamefont
  {Pan}}, \bibinfo {author} {\bibfnamefont {I.}~\bibnamefont {Heemskerk}},
  \bibinfo {author} {\bibfnamefont {C.}~\bibnamefont {Ibar}}, \bibinfo {author}
  {\bibfnamefont {B.~I.}\ \bibnamefont {Shraiman}}, \ and\ \bibinfo {author}
  {\bibfnamefont {K.~D.}\ \bibnamefont {Irvine}},\ }\href@noop {} {\bibfield
  {journal} {\bibinfo  {journal} {Proc. Natl. Acad. Sci. U.S.A.}\ }\textbf
  {\bibinfo {volume} {113}},\ \bibinfo {pages} {E6974} (\bibinfo {year}
  {2016})}\BibitemShut {NoStop}%
\bibitem [{\citenamefont {Ranft}\ \emph {et~al.}(2014)\citenamefont {Ranft},
  \citenamefont {Aliee}, \citenamefont {Prost}, \citenamefont {J{\"u}licher},\
  and\ \citenamefont {Joanny}}]{Ranft2014}%
  \BibitemOpen
  \bibfield  {author} {\bibinfo {author} {\bibfnamefont {J.}~\bibnamefont
  {Ranft}}, \bibinfo {author} {\bibfnamefont {M.}~\bibnamefont {Aliee}},
  \bibinfo {author} {\bibfnamefont {J.}~\bibnamefont {Prost}}, \bibinfo
  {author} {\bibfnamefont {F.}~\bibnamefont {J{\"u}licher}}, \ and\ \bibinfo
  {author} {\bibfnamefont {J.-F.}\ \bibnamefont {Joanny}},\ }\href@noop {}
  {\bibfield  {journal} {\bibinfo  {journal} {New J. Phys.}\ }\textbf {\bibinfo
  {volume} {16}},\ \bibinfo {pages} {035002} (\bibinfo {year}
  {2014})}\BibitemShut {NoStop}%
\bibitem [{\citenamefont {Williamson}\ and\ \citenamefont
  {Salbreux}(2017)}]{Williamson2017}%
  \BibitemOpen
  \bibfield  {author} {\bibinfo {author} {\bibfnamefont {J.~J.}\ \bibnamefont
  {Williamson}}\ and\ \bibinfo {author} {\bibfnamefont {G.}~\bibnamefont
  {Salbreux}},\ }\href@noop {} {} (\bibinfo {year} {2017})\BibitemShut
  {NoStop}%
\bibitem [{\citenamefont {Bissell}\ and\ \citenamefont
  {Hines}(2011)}]{Bissell2011}%
  \BibitemOpen
  \bibfield  {author} {\bibinfo {author} {\bibfnamefont {M.~J.}\ \bibnamefont
  {Bissell}}\ and\ \bibinfo {author} {\bibfnamefont {W.~C.}\ \bibnamefont
  {Hines}},\ }\href@noop {} {\bibfield  {journal} {\bibinfo  {journal} {Nat.
  Med.}\ }\textbf {\bibinfo {volume} {17}},\ \bibinfo {pages} {320} (\bibinfo
  {year} {2011})}\BibitemShut {NoStop}%
\bibitem [{\citenamefont {Harach}\ \emph {et~al.}(1985)\citenamefont {Harach},
  \citenamefont {Franssila},\ and\ \citenamefont {Wasenius}}]{Harach1985}%
  \BibitemOpen
  \bibfield  {author} {\bibinfo {author} {\bibfnamefont {H.~R.}\ \bibnamefont
  {Harach}}, \bibinfo {author} {\bibfnamefont {K.~O.}\ \bibnamefont
  {Franssila}}, \ and\ \bibinfo {author} {\bibfnamefont {V.~M.}\ \bibnamefont
  {Wasenius}},\ }\href@noop {} {\bibfield  {journal} {\bibinfo  {journal}
  {Cancer}\ }\textbf {\bibinfo {volume} {56}},\ \bibinfo {pages} {531}
  (\bibinfo {year} {1985})}\BibitemShut {NoStop}%
\bibitem [{\citenamefont {Weinberg}(2007)}]{weinbergbook}%
  \BibitemOpen
  \bibfield  {author} {\bibinfo {author} {\bibfnamefont {R.~A.}\ \bibnamefont
  {Weinberg}},\ }\href@noop {} {\emph {\bibinfo {title} {The biology of
  cancer}}}\ (\bibinfo  {publisher} {Garland Publishing, Inc.},\ \bibinfo
  {year} {2007})\BibitemShut {NoStop}%
\bibitem [{\citenamefont {Alberts}\ \emph {et~al.}(1994)\citenamefont
  {Alberts}, \citenamefont {Bray}, \citenamefont {Johnson}, \citenamefont
  {Lewis}, \citenamefont {Raff}, \citenamefont {Roberts},\ and\ \citenamefont
  {Watson}}]{molecularbio}%
  \BibitemOpen
  \bibfield  {author} {\bibinfo {author} {\bibnamefont {Alberts}}, \bibinfo
  {author} {\bibnamefont {Bray}}, \bibinfo {author} {\bibnamefont {Johnson}},
  \bibinfo {author} {\bibnamefont {Lewis}}, \bibinfo {author} {\bibnamefont
  {Raff}}, \bibinfo {author} {\bibnamefont {Roberts}}, \ and\ \bibinfo {author}
  {\bibnamefont {Watson}},\ }\href@noop {} {\emph {\bibinfo {title} {Molecular
  Biology of the Cell, 3rd edition}}}\ (\bibinfo  {publisher} {Garland
  Publishing, Inc.},\ \bibinfo {year} {1994})\BibitemShut {NoStop}%
\bibitem [{\citenamefont {{J\"{u}lich Supercomputing Centre}}(2018)}]{jureca}%
  \BibitemOpen
  \bibfield  {author} {\bibinfo {author} {\bibnamefont {{J\"{u}lich
  Supercomputing Centre}}},\ }\href@noop {} {\bibfield  {journal} {\bibinfo
  {journal} {Journal of large-scale research facilities}\ }\textbf {\bibinfo
  {volume} {4}} (\bibinfo {year} {2018})}\BibitemShut {NoStop}%
\end{thebibliography}%
\end{document}